\RequirePackage{fix-cm}
\documentclass[smallextended]{svjour3}       %
\smartqed  %
\usepackage{graphicx}

\usepackage[utf8]{inputenc}
\usepackage[T1]{fontenc}

\usepackage{url}
\usepackage{colortbl}
\usepackage{balance}
\usepackage{xspace}
\usepackage{verbatim}
\usepackage{bold-extra}

\usepackage{listings}
\usepackage{soul}
\usepackage{enumitem}
\usepackage[normalem]{ulem}
\setlist{nolistsep,leftmargin=.5cm}

\usepackage{booktabs}
\usepackage{tabularx}
\usepackage{svg}
\usepackage{calc}
\usepackage{float}

\useunder{\uline}{\ul}{}

\usepackage{caption}
\usepackage{microtype}

\usepackage{amsmath,amsfonts}
\usepackage{algorithmic}

\usepackage{graphicx}
\usepackage{textcomp}

\usepackage{wrapfig}
\usepackage{subcaption}

\usepackage{booktabs,multirow}
\usepackage{changepage,threeparttable}
\usepackage{hhline}

\usepackage{tikz}

\usepackage[dvipsnames]{xcolor}

\usepackage[most]{tcolorbox}

\usepackage{adjustbox}
\usepackage{totcount}
\usepackage{ifthen}

\usepackage{hyperref}
\usepackage[clockwise]{rotating}

\makeatletter
\let\cl@chapter\relax
\makeatother
\usepackage[noabbrev]{cleveref}

\newboolean{showcomments}
\setboolean{showcomments}{true}

\ifthenelse{\boolean{showcomments}}
{
\newcommand{\nb}[2]{
\fbox{\bfseries\sffamily\scriptsize#1}
{\sf\small$\blacktriangleright$\textit{#2}$\blacktriangleleft$}
}
}
{\newcommand{\nb}[2]{}}

\newcommand{\lr}{{\scshape{Logistic Regression}}\xspace}
\newcommand{\gnb}{{\scshape{Na\"{i}ve Bayes}}\xspace}
\newcommand{\knn}{{\scshape{K-Nearest Neighbors}}\xspace}
\newcommand{\svc}{{\scshape{Support Vector Machines}}\xspace}
\newcommand{\dtree}{{\scshape{Decision Trees}}\xspace}
\newcommand{\randforest}{{\scshape{Random Forest}}\xspace}
\newcommand{\rfr}{\randforest}

\newcommand{\lrs}{{\scshape{lr}}\xspace}
\newcommand{\gnbs}{{\scshape{nb}}\xspace}
\newcommand{\knns}{{\scshape{knn}}\xspace}
\newcommand{\svcs}{{\scshape{svm}}\xspace}
\newcommand{\dtrees}{{\scshape{dt}}\xspace}
\newcommand{\randforests}{{\scshape{rf}}\xspace}
\newcommand{\rfrs}{\randforests}

\newcommand{\embgpt}{{\scshape{Gpt$_{emb}$}}\xspace}
\newcommand{\embllama}{{\scshape{Llama$_{emb}$}}\xspace}
\newcommand{\embbert}{{\scshape{Bert$_{emb}$}}\xspace}
\newcommand{\embtfidf}{{\scshape{Tf-Idf}}\xspace}

\newcommand{\llama}{{\scshape{Llama-3}}\xspace}
\newcommand{\llamaft}{{\scshape{Llama-3$_{ft}$}}\xspace}
\newcommand{\gpt}{{\scshape{Gpt-4}}\xspace}
\newcommand{\bert}{{\scshape{Bert}}\xspace}
\newcommand{\dbert}{{\scshape{DistilBert}}\xspace}
\newcommand{\rbert}{{\scshape{RoBERTa}}\xspace}
\newcommand{\xlnet}{{\scshape{XLNet}}\xspace}

\newcommand\rev[1]{{\color{black}{#1}}}

\newcommand{\ie}{\textit{i.e.},\xspace}
\newcommand{\eg}{\textit{e.g.},\xspace}
\newcommand{\etc}{\textit{etc.}\xspace}
\newcommand{\etal}{\textit{et al.}\xspace}
\newcommand{\aka}{\textit{a.k.a.}\xspace}

\definecolor{bug_red}{rgb}{.84,.23,.29}

\definecolor{info-needed-color}{rgb}{1,.8,.12}

\newboolean{showboxes}
\setboolean{showboxes}{true}

\newcounter{findingcounter}
\newtotcounter{totalfindings}

\ifthenelse{\boolean{showboxes}}{

\newcommand{\rqanswer}[1]{
\begin{tcolorbox}[enhanced,skin=enhancedmiddle,
borderline={1mm}{0mm}{MidnightBlue},
boxsep=3pt]
\normalsize #1
\end{tcolorbox}
\addtocounter{totalfindings}{1}
}

}{

\newcommand{\rqanswer}[1]{}
}

\begin{document}

\title{Evaluating Language Model Applications for Identifying Solution-Related Content in Issue Report Discussions%
}

\author{Antu Saha        \and
        Mehedi Sun \and Oscar Chaparro
}

\institute{A. Saha, M. Sun, and O. Chaparro \at
              Department of Computer Science \\
              William \& Mary\\
              Williamsburg, VA, USA\\
              \email{\{asaha02, msun12, oscarch\}@wm.edu}           
}

\date{Received: date / Accepted: date}

\maketitle

\begin{abstract}
During issue resolution, software developers rely on issue reports to discuss solutions for defects, feature requests, and other changes. These discussions contain proposed solutions—from design changes to code implementations—as well as their evaluations. Locating solution-related content is essential for investigating reopened issues, addressing regressions, reusing solutions, and understanding code change rationale. Manually understanding long discussions to identify such content can be difficult and time-consuming.

This paper automates solution identification using language models as supervised classifiers. We investigate three applications—embeddings, prompting, and fine-tuning—across three classifier types: traditional ML models (MLMs), pre-trained language models (PLMs), and large language models (LLMs). Using 356 Mozilla Firefox issues, we created a dataset to train and evaluate six MLMs, four PLMs, and two LLMs across 68 configurations.

Results show that MLMs with LLM embeddings outperform \embtfidf features, prompting underperforms, and fine-tuned LLMs achieve the highest performance, with \llamaft reaching 0.716 F1 score. Ensembles of the best models further improve results (0.737 F1). Misclassifications often arise from misleading clues or missing context, highlighting the need for context-aware classifiers. Models trained on Mozilla transfer to other projects, with a small amount of project-specific data, further enhancing results. This work supports software maintenance, issue understanding, and solution reuse.

\end{abstract}

\section{Introduction}
\label{sec:intro}
During the evolution and maintenance of a software project, developers are tasked with resolving software \textit{issues}, including defects, new features, enhancements, and other change requests~\cite{rajlich2011software,zeller2009programs}. 
To address these issues, developers typically engage in a range of activities, such as issue triage, analysis, solution design, implementation, and the verification and validation of such implementation~\cite{zhang2016literature,saha2015understanding,zeller2009programs,zou2018practitioners}.
This process, known as issue resolution~\cite{zhang2016literature,saha2015understanding,zeller2009programs,zou2018practitioners}, is typically iterative, incremental, and collaborative. It often involves discussions among multiple stakeholders (\eg reporters, developers, project managers, Q/A members, \etc) to collectively and incrementally design, implement, and assess solutions to the issues~\cite{ko2011design}.

These discussions are frequently captured in issue reports through issue tracking systems (\eg GitHub Issues~\cite{github}, Jira~\cite{jira}, or Bugzilla~\cite{bugzillaBugzilla}), where developers contribute comments expressing ideas, opinions, and information relevant to issue resolution~\cite{Zimmermann2009,Zimmermann2010,ko2011design,rath2020request}. Specifically, developers propose solutions discussed, evaluated, and refined before a final solution is implemented~\cite{brunet2014developers,ko2011design}. In this paper, we refer to \textit{solution-related content} as the textual information within issue report comments that capture these proposed solutions, their assessments, and the implementation details.
\looseness=-1

Locating solution-related content within issue report comments is highly important for developers as it serves multiple purposes. For instance, when addressing regressions or reopened issues, developers need to locate and review this content in past issue reports to understand how the solutions were derived, their impact, and the reasons behind new or recurring problems~\cite{Zimmermann2012,shihab2010predicting}. Additionally, when understanding code~\cite{tao2012software}, developers can consult solution-related content to gain insights into the rationale behind code design and implementations~\cite{latoza2010hard,sillito2006questions,brunet2014developers,kleebaum2021continuous}, providing context for new and recurring problems and documenting the evolution of prior implemented solutions. Solution-related content can also guide the reuse and design of solutions for new problems that resemble or are related to previously resolved issues~\cite{Zimmermann2010,bettenburg2008duplicate,li2018issue}. In open-source projects, this content plays a key role in making decisions on various aspects of the project (\eg its overall evolution~\cite{tsay2014let}) and helps newcomers learn how solutions are developed and decided upon within the project~\cite{gilmer2023summit}.
\looseness=-1

Despite its importance, manually identifying solution-related content in issue report discussions is often a time-consuming and challenging task. These discussions can include tens or even hundreds of interactions (in the form of comments or posts) among various contributors~\cite{saha2025decoding,gilmer2023summit,ko2011design}, addressing different aspects of issue resolution, which makes following the conversation and extracting relevant information difficult~\cite{arya2019analysis,huang2018automating,gilmer2023summit}. Previous studies~\cite{saha2025decoding,gilmer2023summit,ko2011design} have shown that understanding issue report discussions demands significant mental effort from developers, as the conversations can be entangled and redundant, containing various types of information and overlapping content threads that are difficult to follow.  This problem is further compounded by the large volume of issue reports and comments developers must process, a common scenario in many software projects~\cite{zou2018practitioners,anvik2006should}.

\rev{Identifying solution-related content in issue report discussions can be framed as a text classification task, where various ML and DL models serve as binary classifiers to determine if a given text passage (\eg a sentence or a paragraph) conveys information of a certain kind (\eg a proposed solution to the issue).} However, it remains unclear how different types of models—namely, machine learning models (MLMs), pretrained language models (PLMs), and large language models (LLMs)—perform on this task. In addition, the potential of combining traditional MLM-based classifiers (\eg\ \svc) with LLM-based embeddings has yet to be explored. Finally, it remains unclear whether integrating different types of models can further improve classification performance.

This paper addresses these research gaps by conducting an empirical study that leverages MLMs, PLMs, and LLMs techniques to automatically identify solution-related content in developers’ discussions. Focusing on Mozilla Firefox, a well-established open-source project, we examined three applications of language models—\textit{embeddings}, \textit{prompting}, and \textit{fine-tuning}—across three types of classifiers: MLMs, PLMs, and LLMs. Using a dataset from a recent work~\cite{saha2025decoding} (comprising 788 solution-related and 4,079 non-solution-related comments from 356 Mozilla Firefox issue reports), we trained and evaluated six MLMs, four PLMs, and two LLMs across 68 configurations.

\looseness=-1

Our study provides key insights into the effectiveness of different classifier families for identifying solution-related comments. Models using language model embeddings consistently outperform traditional \embtfidf encodings, highlighting the advantages of semantically richer text representations for this task. Among lightweight classifiers, \svc paired with {\scshape{Gpt}}  embeddings achieves strong performance (F1 = 0.663), demonstrating that teams with limited computational resources can still benefit from pretrained embeddings without requiring large-scale models. However, the highest performance comes from task-specific fine-tuning. The fine-tuned \llama (\ie~\llamaft) and \rbert models achieve F1 scores of 0.716 and 0.692, respectively, substantially outperforming both traditional MLMs and prompted LLMs. These gains show that fine-tuning enables models to learn the linguistic and contextual patterns of solution discussions that are not easily captured through prompting alone. Qualitative analysis further indicates that misclassifications often stem from misleading clues (\eg keywords or code snippets) or lack contextual information (\eg issue description and surrounding discussion), suggesting that future approaches should incorporate richer discourse- or thread-level representations rather than analyzing comments in isolation.

We also experimented with ensemble strategies that combine different model types. The ensemble of the best-performing classifiers—\svc, \rbert, and \llamaft—achieves an F1 score of 0.737, improving on the best single model (\ie\ \llamaft) by 2.9\%. This result emphasizes that different model families capture distinct markers of solution-related reasoning, and leveraging these complementary strengths leads to more reliable identification. Additionally, our cross-project evaluation shows that knowledge learned from the Mozilla dataset transfers effectively to other projects such as Chromium and GnuCash. Performance improves further when models are supplemented with only a small amount of project-specific data, demonstrating that organizations can adopt these models with minimal annotation overhead. This offers a practical deployment strategy: teams can begin with our pretrained models and incrementally refine them as they encounter new issue discussions, supporting scalable and maintainable integration of automated solution identification into real-world workflows.

In summary, we make the following main contributions:
\begin{enumerate}
    \item A comprehensive study that assesses the effectiveness of 12 text classifiers of different kinds that leverage language models in identifying solution-related content in Mozilla Firefox issue report discussions.
    
    \item An analysis of the effectiveness of combining individual classifiers of different kinds to enhance the identification of solution-related content.

    \item A qualitative analysis of the predictions by the three best-performing models of each category (\ie\ \llamaft, \rbert, and \svcs) to gain deeper insights into the models’ decision-making behavior.
    
    \item A generalizability study on Chromium~\cite{chromium} and GnuCash~\cite{gnucash} projects to demonstrate if the knowledge learned by the models from the Mozilla dataset can transfer to other projects.

    \item A complete replication package with the source code, data, and documentation for replicating and verifying our experiments and results~\cite{repl_pack}.
\end{enumerate}

\section{Problem, Motivating Example, and Related Work}
\label{sec:related_work}

\subsection{Problem and Motivating Example}
This paper aims to automatically identify \textit{solution-related content} in developers’ discussions in issue reports. We formulate this problem as a binary text classification task, where a classifier takes issue report comments as input and predicts whether each comment is \textit{solution-related} or \textit{non-solution-related}. As the classifier, we can leverage different types of models, \eg\ MLMs, PLMs, or LLMs.

\subsubsection{Complexity of issue report discussions}
A few studies have characterized the dynamics and complexities of issue discussions and highlighted the need for automated support to help developers navigate and understand these conversations. Ko \etal~\cite{ko2011design} analyzed issue report discussions and found that they are often centered around achieving the desired design intent or adapting to user needs, with decision-making dominated by project leaders and based on anecdotes, speculation, and generalizations. These factors contribute to the complexity of issue discussions, where important design and solution-related information is frequently buried among numerous comments, making it difficult for stakeholders to follow the conversation and extract the relevant content.  Similarly, Gilmer \etal~\cite{gilmer2023summit} explored the characteristics of content summaries within issue discussions and identified the challenges faced by developers when reviewing and contributing to these threads. They found that issue discussions are often entangled and confusing, prompting developers to adopt strategies like skimming or relying on visual signals (\eg comment reactions) to locate relevant information. Both studies highlight the need for automated content classifiers or summarization tools to reduce the cognitive load on developers and improve the efficiency of understanding issue report discussions.
\looseness=-1

\subsubsection{Motivating Example}
\begin{figure}[t]
	\centering
	\includegraphics[width=1\linewidth]{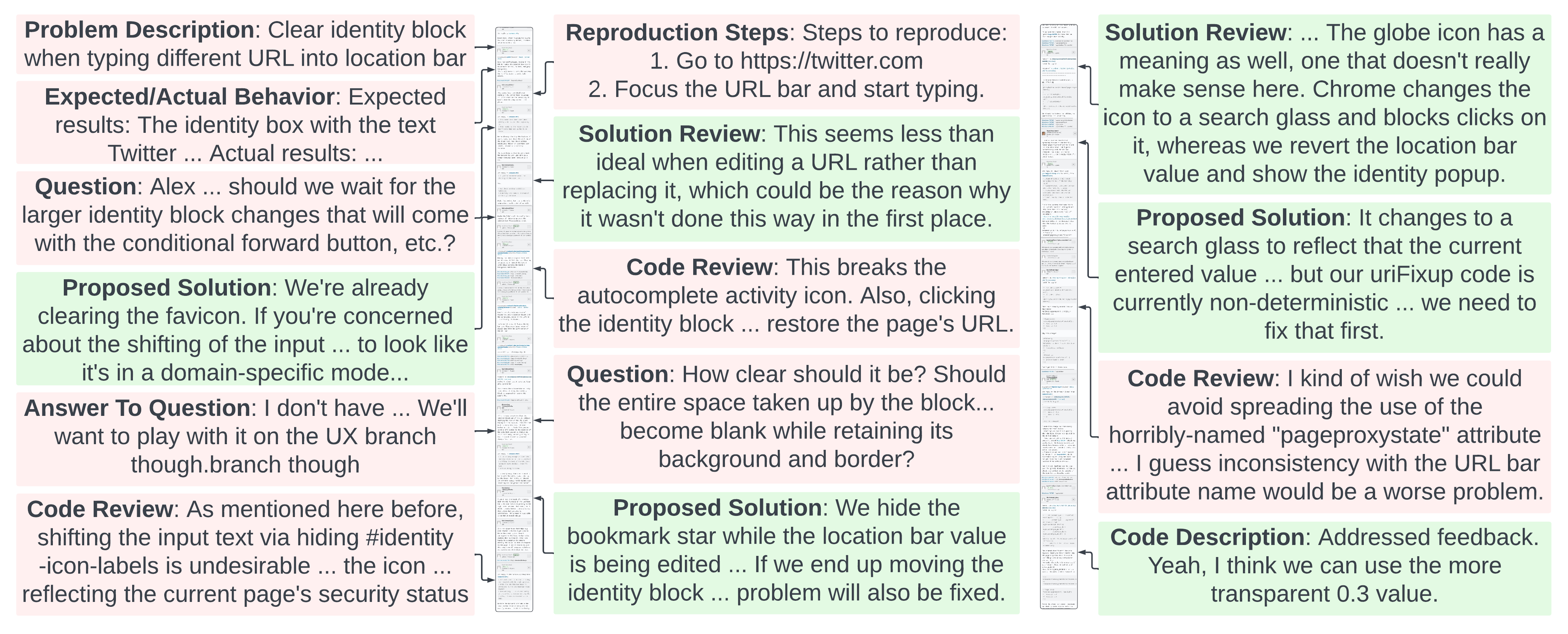}
	\caption{Firefox's Issue Report \#667586~\cite{bug}, split into two parts, which contain 36 comments with different kinds of information. Solution-related comments are displayed in the green boxes.}
	\label{fig:issue_example}
\end{figure}

We illustrate the complexity of issue discussions with issue report \#667586~\cite{bug} from Mozilla Firefox, the subject of our case study. The issue report, visualized in \Cref{fig:issue_example}, reports a defect titled "Clear identity block when typing a different URL into the location bar". 
\looseness=-1

This issue included an extensive discussion among 13 stakeholders (mainly developers), resulting in 37 comments. These comments, comprising 69 paragraphs and 122 sentences, covered a wide range of information types. Notably, they contained \textit{solution-related content} (highlighted in green in \cref{fig:issue_example}), which includes discussions about \textit{proposed solutions}, \textit{solution reviews}, and/or \textit{code descriptions} of the implemented solution. In addition, the issue includes problem descriptions, expected and observed system behavior, reproduction steps, questions, clarifications, and code reviews. 
These contributions resulted from efforts to triage, reproduce, and investigate the issue, as well as to design, implement, and validate a solution.
\looseness=-1

Resolving this defect required multiple iterations and collaboration among stakeholders. After several failed attempts to solve the issue, it remained inactive for nearly a year before a new developer was assigned to it. To resolve the issue, this developer needed to read through the lengthy conversation to understand previous efforts by other team members. The intertwining of solution-related comments with other discussions likely made it difficult for the developer to comprehend the solutions attempted earlier.

This example highlights the need for automated methods to help developers quickly locate solution-related content. In this paper, we evaluate text classification approaches based on language models to identify such content. 
\rev{While in this work we focus on evaluating these models, with some engineering effort, they can be integrated into issue trackers as practical tools to automatically identify and visually label solution-related comments in issue discussions, so that developers can easily identify such content when inspecting the discussions.}

\subsection{Related Work}

Extensive research has explored the use of supervised models for text classification to identify fine-grained snippets that convey different information types in natural language software artifacts (\ie sentences, paragraphs, or comments/posts, as opposed to entire artifacts~\cite{brunet2014developers,pan2021automating,shang2024analyzing,colavito2024leveraging,catolino2019not}). These artifacts include code reviews~\cite{zanaty2018empirical}, pull requests~\cite{viviani2019locating}, email threads~\cite{fu2021machine,di2015development,bacchelli2012content}, developer forums~\cite{uddin2019automatic}, chat conversations~\cite{shang2024analyzing,pan2021automating}, commits~\cite{shakiba2016fourd}, code comments~\cite{da2017using,rani2021identify}, app reviews~\cite{Panichella2015}, requirement documents~\cite{luo2022prcbert}, and issue reports~\cite{wang2020argulens,arya2019analysis,huang2018automating,brunet2014developers,song2020bee,chaparro2019assessing,zhao2019automatically,adnan2025sprint}. 

In the context of issue reports, prior work has focused on identifying developer intentions (\eg giving or seeking information or proposing a new feature)~\cite{huang2018automating}, bug description elements (\eg bug reproduction steps or the observed/expected system behavior)~\cite{mahmud2025combining,song2020bee,chaparro2019assessing,zhao2019automatically}, buggy UI screens and components~\cite{saha2024toward}, design-related content (\eg code structure and design decisions)~\cite{brunet2014developers,zhao2024drminer}, argument elements~\cite{wang2020argulens}, technical debt~\cite{skryseth2023technical}, bug resolution-related comments~\cite{krasniqi2021recommending}, and other information types (\eg issue investigation and exploration or task progress)~\cite{arya2019analysis,mehder2022classification}. Beyond issue reports, studies on other software artifacts have identified content related to software architecture and design (\eg rationale and decisions)~\cite{zanaty2018empirical,shakiba2016fourd,viviani2019locating}, project decision-making~\cite{fu2021machine}, issue-solution pair~\cite{shi2021ispy}, self-admitted technical debt~\cite{gu2024self}, API quality aspects~\cite{uddin2019automatic}, requirement types~\cite{luo2022prcbert}, and more broadly, intentions, topics, or various other information types~\cite{di2015development,bacchelli2012content,rani2021identify,pan2021automating,shang2024analyzing}.

We build on this body of work by specifically focusing on detecting solution-related content in issue report discussions via language models. This content includes solution descriptions for different types of issues (\eg defects and enhancements) and problems (\eg UI issues), assessment on these solutions, and details on how final solutions were refined and implemented.

Most classification approaches in the field employ traditional machine learning models (MLMs), such as SVMs or Random Forests,  trained and evaluated on domain-specific datasets using features learned from the text (\eg words and n-grams encoded via TF-IDF, GloVe, or Word2Vec) or hand-crafted features extracted from different sources (\eg part-of-speech tags, linguistic patterns, or content authorship)~\cite{wang2020argulens,fu2021machine,arya2019analysis,brunet2014developers,Panichella2015,zanaty2018empirical,di2015development,bacchelli2012content,uddin2019automatic,shakiba2016fourd,da2017using,rani2021identify,song2020bee,zhao2019automatically}. To the best of our knowledge, we are the first to investigate the use of MLMs leveraging pre-trained language model embeddings to identify solution-related content in issue report discussions.

Deep learning (DL) approaches for text classification have also been studied due to their ability to capture richer text semantics. These include models like LSTMs and CNNs~\cite{pan2021automating,huang2018automating,chaparro2019assessing,shi2021ispy}, and more recently, Transformer-based language models such as BERT~\cite{shang2024analyzing,colavito2024leveraging,luo2022prcbert,zhao2024drminer,mehder2022classification}, which have demonstrated superior performance over earlier architectures~\cite{chang2024survey,li2022survey}. DL models are usually pre-trained on large, general-purpose corpora (\eg news or Wikipedia articles) to learn semantic representations of text, and are used for downstream tasks like text generation or classification~\cite{chang2024survey,li2022survey}. Although DL models can be employed in zero-shot or few-shot settings (\eg via prompting), fine-tuning them on domain-specific datasets typically yields better results (as we show in our study)~\cite{chang2024survey,li2022survey,minaee2024large}. While the use of language models have been employed for diverse generative tasks in SE (code generation, summarization, and review~\cite{shin2023prompt,wang2022no,pornprasit2024gpt} and other tasks~\cite{hou2023large,watson2022systematic}), our contribution lies in the empirical evaluation of prompting and fine-tuning pre-trained transformer-based LMs (PLMs) and large LMs (LLMs) for identifying solution-related content in issue discussions, which has not yet explored in previous research (to the best of our knowledge).

Lastly, while prior studies have combined models within the same group— either traditional ML or DL models—for various tasks~\cite{huang2018automating,fu2021machine}, we explore integrating predictions across multiple model groups—MLMs, PLMs, and LLMs—to more accurately identify solution-related content in developers’ discussions.

\section{Study Methodology}
\label{sec:methodology}

\subsection{Research Goal and Questions}
This study aims to evaluate the effectiveness of supervised text classifiers based on language models (LMs) in identifying solution-related comments in issue report discussions (\aka solution identification). The study focuses on issue discussions from Mozilla Firefox~\cite{mozilla-firefox}. We selected  Firefox as the subject of the study for three main reasons: (1) it is an active and widely used open-source project with more than 19 years of evolution, and (2) it extensively uses its issue tracker for discussing and managing project issues, including defects, new features, enhancements, and tasks~\cite{firefox-bug-handling}, and (3) we studied this project in our prior work that investigated Mozilla's issue resolution process~\cite{saha2025decoding}.
\looseness=-1

The study evaluates three classifier types: traditional machine learning models  (MLMs), pre-trained LMs (PLMs), and Large LMs (LLMs), across three LM applications:  \textit{embeddings}, \textit{prompting}, and \textit{fine-tuning}. In addition, the study evaluates combinations of these models to assess if they achieve superior classification performance. The study aims to answer the following research questions (RQs):
\looseness=-1

\begin{enumerate}[label=\textbf{RQ$_\arabic*$:}, ref=\textbf{RQ$_\arabic*$}, itemindent=0cm,leftmargin=1cm]
	\item \label{rq:embed}{How effective are ML models with \textit{LM embeddings} at identifying solution content?}
	\item \label{rq:prompt}{How effective are \textit{prompted} language models at identifying solution content?}
	\item \label{rq:finetune}{How effective are \textit{fine-tuned} language models at identifying solution content?}
	\item \label{rq:ensemble}{How effective are \textit{combinations of ML and language models}  at identifying solution content?}
    \item \label{rq:types}{How do models' performance vary across different \textit{issue types} and \textit{problem categories}?}
	
\end{enumerate}

\begin{figure}[t]
	\centering
	\includegraphics[width=1\linewidth]{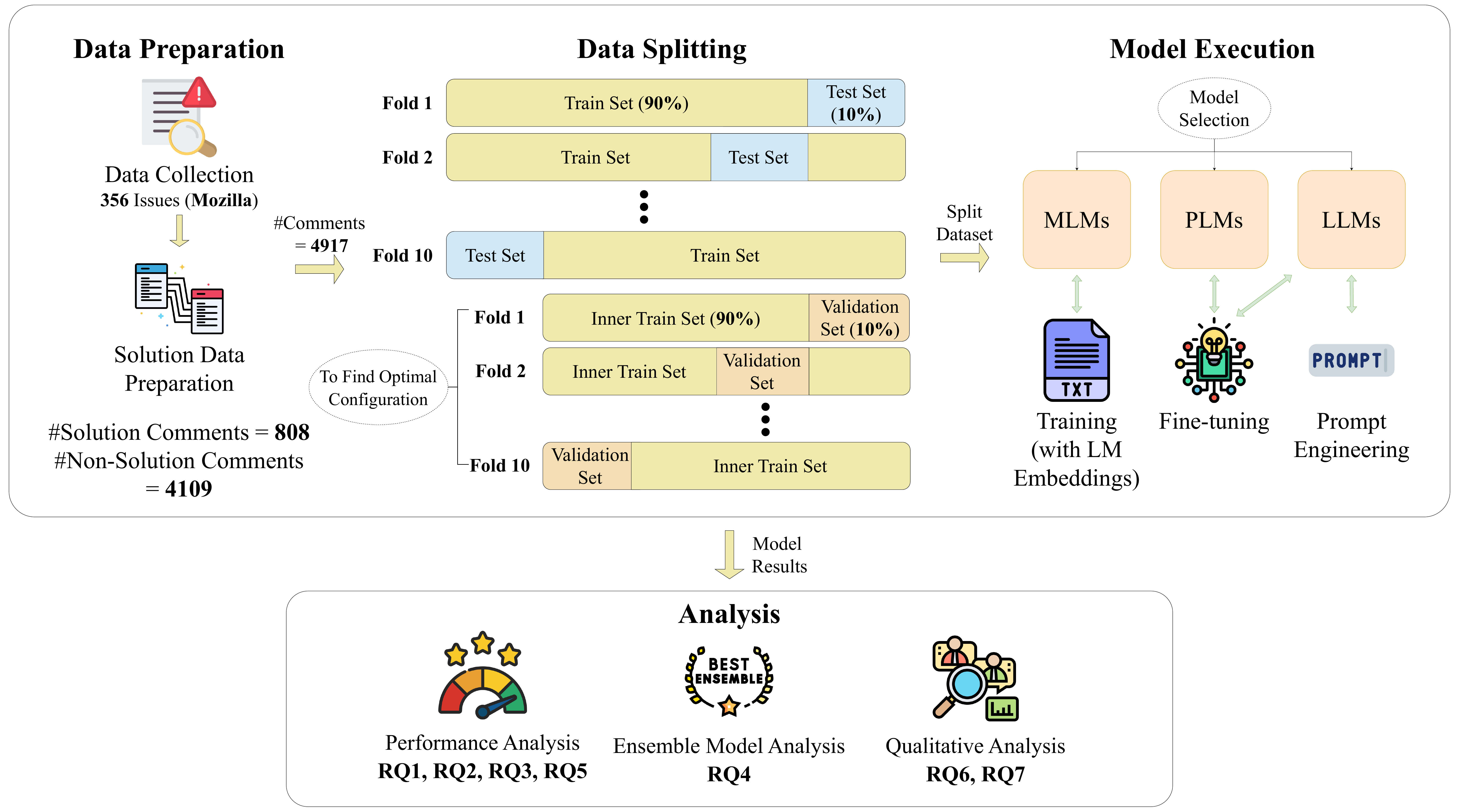}
	\caption{Study Methodology}
	\label{fig:research_methodology}
\end{figure}

\ref{rq:embed} investigates the effectiveness of MLMs (\eg\ \svc or \rfr) when issue content is encoded with various LMs (\eg\ \bert or \llama). 
\ref{rq:prompt} investigates the effectiveness of LLMs (\ie\ \llama) when prompted for classification of issue solution content\footnote{We experiment with LLMs and not with PLMs as the former have shown superior performance in many downstream generative tasks (\eg Q\&A)~\cite{hou2023large}.}. \ref{rq:finetune} investigates the effectiveness of classifiers based on PLMs (\eg\ \bert) and LLMs (\llama) fine-tuned for solution identification. \ref{rq:ensemble} investigates whether combining classifiers of different types (\eg MLMs and PLMs) leads to better classification performance. \ref{rq:types} investigates how model performance varies across different issue types (\eg defects, enhancements, tasks) and problem categories (\eg UI issues, crashes, code improvements). \Cref{fig:research_methodology} shows an overview of the methodology we employed to answer the RQs.
\looseness=-1

Moreover, we conducted a qualitative analysis of model predictions to gain deeper insights into their decision-making behavior (see \Cref{sec:qualitative_analysis}) and a generalizability study using the Chromium~\cite{chromium} and GnuCash~\cite{gnucash} projects to assess whether the knowledge learned from Mozilla data transfers effectively to other projects (see \Cref{sec:generalizability}).

\subsection{Dataset Collection}
\label{sub:data_col}

We utilized the dataset constructed in our prior work on understanding the issue resolution process adopted in Mozilla Firefox~\cite{saha2025decoding}. This dataset comprises 356 thoroughly annotated issues, where each issue report comment was manually labeled based on the information it conveys. The annotation process in our prior work was rigorous and carefully designed, following a multi-coder, iterative \textit{open coding} methodology~\cite{spencer2009card} involving seven trained researchers with both academic and industry experience. 

\rev{We provide a summary of the annotation procedure in this paper, but refer the reader to the original paper for more details~\cite{saha2025decoding}. Each issue report was independently annotated and validated by at least two researchers, and all annotations were guided by a collaboratively developed code catalog and detailed coding guidelines to ensure consistency and reliability. During annotation, researchers examined the textual content of issue comments and labeled relevant text snippets (\ie phrases, sentences or paragraphs) that describe activities performed during issue resolution as well as the types of problems reported in the issues. The process was conducted iteratively in batches of issues. In each batch, a first annotator assigned codes to relevant text snippets in the comments, and a second annotator reviewed the annotations to verify their accuracy and completeness. Disagreements were resolved through discussion and reconciliation sessions, and when necessary, a third annotator helped resolve conflicts. Throughout the process, the code catalog and coding guidelines were continuously refined to ensure consistent interpretation and application of the codes across annotators (see~\cite{saha2025decoding} for more details).}

Mozilla Firefox is a multi-language, multi-platform project with well-documented issue management practices~\cite{firefox-bug-handling}. The project uses BMO, a modified version of the Bugzilla issue tracker~\cite{bugzillaBugzilla}, as its primary platform for managing system change requests (\ie issues)~\cite{firefox-bug-pipeline}. Three types of issue reports can be created via BMO: defects, enhancements, and tasks~\cite{mozilla-products}. While Mozilla uses BMO to manage issues for both Firefox Desktop and Firefox Mobile~\cite{mozilla-products}, our dataset includes only the issues for Firefox Desktop, covering its two main components: \textit{Firefox} and \textit{Core}. The \textit{Firefox} component implements the browser’s graphical user interface (GUI), while the \textit{Core} component provides essential functionalities such as web browsing, rendering, and networking services.

The dataset includes only \textit{fixed} and \textit{resolved} issue reports from the selected components. In our prior study, we used Bugzilla’s API~\cite{bugzilla-api} to collect all issues created between January 1, 2010, and April 30, 2023, thereby covering a long period of system evolution. The downloaded data contained issue titles/summaries, descriptions, comments, and relevant metadata such as creation and resolution times. From the 199,271 collected issues (approximately 164.7k for \textit{Core} and 34.5k for \textit{Firefox}), we selected a statistically significant sample of 384 issues with a 95\% confidence level and a 5\% error margin. After discarding 28 automatically generated pull request issues, the final dataset contained \textbf{356 issues}.

\begin{table}[h]
\centering
\caption{Examples of Annotation Codes}
\label{tab:annotation_codes}
\resizebox{\columnwidth}{!}{%
\begin{tabular}{>{\centering\arraybackslash}m{3.5cm} 
                >{\centering\arraybackslash}m{2.5cm} 
                >{\centering\arraybackslash}m{2.5cm} 
                >{\centering\arraybackslash}m{8cm}}
\toprule
\textbf{Code} & \textbf{\# of Annotations} & \textbf{\# of Issues} & \textbf{Real Example} \\
\midrule
\textbf{\texttt{CODE\_DESCRIPTION}} & 885 & 328 & OK, fixing things turned out to be a bit more complicated than a simple backout... \\
\midrule
\textbf{\texttt{CODE\_REVIEW}} & 1228 & 264 & We already null-check for gURLBar, so remove this \texttt{if} condition. \\
\midrule
\textbf{\texttt{PROPOSED\_SOLUTION}} & 256 & 142 & If it is easy enough to clear the identity block out of the UI, without shifting the text of the URL (just having it turn invisible)... \\
\midrule
\textbf{\texttt{SOLUTION\_VERIFICATION}} & 197 & 115 & I guess that's not really feasible, given the conversion from relative units to computed values depend on this too... \\
\midrule
\textbf{\texttt{PROBLEM\_CAUSE}} & 123 & 76 & Every Intel card has been declared as “WebGL not recommended,” so... \\
\midrule
\textbf{\texttt{PROBLEM\_REVIEW}} & 145 & 74 & Tried ‘baseline’ and it turned out that the position is not centered but slightly... \\
\midrule
\textbf{\texttt{SOLUTION\_REVIEW}} & 175 & 64 & I guess that's not really feasible, given the conversion from relative units to computed values depend on this too... \\
\midrule
\textbf{\texttt{PROBLEM\_REVIEW\_REQUEST}} & 64 & 47 & Sotaro: could you take a look at this? Thanks. \\
\midrule
\textbf{\texttt{REPRODUCTION\_ATTEMPT}} & 127 & 47 & I tested this on my HP TouchSmart tx2 laptop which runs Windows 7 and that machine loads the sites... \\
\midrule
\textbf{\texttt{NEW\_ISSUE\_FILING}} & 76 & 46 & Filed bug 687113 for simplifying that stuff. Someone take it—it's a nice opportunity to improve behavior and delete code at the same time! \\
\bottomrule
\end{tabular}%
}
\end{table}

\begin{table}[h]
\centering
\caption{Example of Problem Categories}
\label{tab:prob_categories}
\resizebox{.8\columnwidth}{!}{%
\begin{tabular}{>{\centering\arraybackslash}m{3.5cm} >{\centering\arraybackslash}m{2cm} >{\centering\arraybackslash}m{8cm}}
\toprule
\textbf{Problem Category} & \textbf{\# of Issues} & \textbf{Real Examples} \\
\midrule
\textbf{UI Issue} & 33 & Incorrect alignment of a label; Scrollbar is missing. \\
\midrule
\textbf{Crash} & 23 & Firefox crashes when closing the window in fullscreen; software crashes due to assertion failure. \\
\midrule
\textbf{Technology Update} & 8 & Update library; Necessary components are not installed. \\
\midrule
\textbf{Error Handling Improvement} & 7 & Show specific error messages; Add warning if someone runs shutdown CC more than twice. \\
\midrule
\textbf{Preventive Changes} & 6 & Compiling the SIPCC code with log macros expanded directly to \texttt{printf()} yields about 300 formatting-related warnings. \\
\midrule
\textbf{Incorrect Page Rendering} & 8 & Content from AMO is restricted by \texttt{X-FRAME-OPTIONS} header; Crash removing full-screen element during “resize” event; Dark mode colors set too late in activity stream startup. \\
\midrule
\textbf{Audit Task} & 1 & Audit the implementation of a function. \\
\midrule
\textbf{Compatibility Issue} & 4 & WebGL not working on Windows XP; Skia unable to compile MacPPC and Android-x86. \\
\midrule
\textbf{Code Improvement} & 32 & Make explicit constructors from implicit constructors; Fix warnings; Remove dependency. \\
\midrule
\textbf{Unnecessary Code Removal} & 19 & Remove disable-mathml support. \\
\bottomrule
\end{tabular}%
}
\end{table}

In total, these 356 issue reports included 4,917 manually annotated comments. The annotation process resulted in \textbf{22 annotation codes} representing different types of information discussed in the comments and \textbf{17 problem categories} representing the types of software problems. \Cref{tab:annotation_codes,tab:prob_categories} presents examples of these codes and problem categories, respectively, while the complete catalog with definitions, rules, and examples is available in our replication package~\cite{repl_pack}. 

In this paper, we leveraged the annotated dataset from our prior work for two primary purposes. First, we used the annotation codes to identify and label comments discussing solutions, thereby constructing the ground-truth dataset for our solution classification models (see \Cref{sub:data_preparation}). Second, we employed the problem categories associated with each issue to analyze and interpret the models’ classification performance across diverse problem domains (see \Cref{sec:results_across_prob_cat}).

\subsection{Solution Data Preparation}
\label{sub:data_preparation}

\rev{In this study, we defined an issue comment as a \textit{solution comment} if it discusses any proposed solution to the reported issue, assesses a proposed solution, or describes the changes made to the code to implement a proposed solution. While solution-related discussions can take different forms (\eg proposing an approach, evaluating an alternative, or describing an implemented fix), these activities all correspond to a common objective—designing or implementing a solution to the reported problem. Therefore, we grouped them into a single \textit{solution comment} class to distinguish comments that contribute to the issue’s resolution from those that discuss other aspects of the issue (\eg reproduction attempts, clarification questions, or code review requests).

To operationalize this definition, we qualitatively examined all 22 annotation codes along with their definitions, rules, and representative examples, and identified those that aligned with our criteria for \textit{solution comment}. For instance, a \textit{solution comment} may describe the strategies, algorithms, or approaches that developers proposed to solve the problem or complete the task reported in the issue (\eg the \texttt{PROPOSED\_SOLUTION} code in \Cref{tab:annotation_codes}). Comments annotated with such codes were labeled as \textit{solution comments}, while all remaining comments were categorized as \textit{non-solution comments}.

Because the original dataset was annotated at the text-snippet level, each comment was carefully examined to ensure consistency between snippet-level annotations and comment-level labels. If a comment contained at least one snippet annotated with a solution-related code, the comment was labeled as a \textit{solution comment}. Conversely, comments without any solution-related codes were labeled as \textit{non-solution comments}. In cases where comments included both solution-related and other types of information, the presence of a solution-related snippet determined the label, while the remaining text was treated as contextual information, which may help a model identify solution content. Our manual inspection of the code catalog and annotated examples ensured that comments discussing solutions were consistently captured by the corresponding codes.}

Of the 4,917 annotated comments, 808 (16.5\%) were labeled as \textit{solution comments}, while the remaining 4,109 (83.6\%) are \textit{non-solution comments}. These labeled comments serve as the ground truth dataset for the classification models.
On average, each issue includes 11.5 non-solution comments with 2.3 solution comments, and 15.6\% of the comments in an issue report are solution comments.
We used this dataset for evaluating the classification models. We focused on entire comments rather than sentences or paragraphs, as comments provide more comprehensive context for users of a solution identification model. Sentences and paragraphs may lack sufficient context; for instance,  a solution description may span multiple related sentences or paragraphs that need to be considered together for a full understanding of the content.
\looseness=-1

\subsection{Model Selection}
\label{sub:approaches}

\subsubsection{Machine Learning Models}

We selected six machine learning models of different kinds, commonly used in prior studies~\cite{arya2019analysis,zanaty2018empirical,shakiba2016fourd,li2020automatic}, for binary classification of issue comments:

\begin{itemize}
    \item \lr (\lrs)~\cite{palanivinayagam2023twenty} is a linear model that learns the relationship between the input features and the log-odds of a comment being a solution or a non-solution. It assigns weights to each feature and predicts the probability of a class, classifying based on a threshold (\eg 0.5).

    \item \knn (\knns)~\cite{peterson2009k} is a non-parametric model that classifies a comment by comparing it to the 
    \textit{k} most similar instances in the training data, using a distance metric (\eg Euclidean or cosine similarity). The class is determined by a majority vote of the nearest \textit{k} neighbors of the input comment. 

    \item \gnb (\gnbs)~\cite{webb2010naive} is a probabilistic classifier based on Bayes' Theorem, assuming independence between features. It calculates the probability of each class given the input features and selects the class with the highest probability. When dealing with continuous features, \gnbs assumes that each feature follows a Gaussian (\ie normal) distribution within each class.

    \item \svc (\svcs)~\cite{hearst1998support} finds a hyperplane that best separates solution and non-solution classes in the feature space. It maximizes the margin between the close points of the classes and can handle both linear and non-linear classification via kernel functions.

    \item \dtree (\dtrees)~\cite{de2013decision} is a tree-like model that splits the feature space into subspaces by choosing the feature that best separates the data at each step. The process continues until leaves are reached, representing the final class labels (solution or non-solution).
    
     \item \rfr (\rfrs)~\cite{rigatti2017random} builds multiple decision trees using random subsets of the data and features. The final prediction is made based on the majority vote from the individual trees.
     
\end{itemize}

\rev{These models were selected to represent a diverse set of classical classification paradigms, including linear models (\eg\ \lr), distance-based learning (\eg\ \knn), probabilistic models (\eg\ \gnb), margin-based classifiers (\eg\ \svc), and tree-based learners (\eg\ \dtree and \rfr). Such models have been widely used in software engineering research given their advantages: they are computationally efficient to develop and use, and have been shown to be effective for textual classification in software artifacts (see \Cref{sec:related_work}). Given that our goal is to evaluate how well modern language model representations capture solution-related contents in issue discussions, these lightweight classifiers provide a controlled and practical setting for assessing the predictive power of different embeddings without introducing additional architectural complexity.}

\subsubsection{Language Models}

We selected six state-of-the-art Transformer-based language models (LMs), which have been used in prior studies for various software engineering tasks~\cite{zhang2023survey,hou2023large}. Of these six, two are large language models (LLMs) with architectures exceeding a billion parameters, while the remaining are pre-trained language models (PLMs) with model sizes in the millions of parameters. 

In our study, due to computational, practical, and cost constraints, three LMs (\ie\ \gpt, \llama, and \bert) are used to generate embeddings for the machine learning-based solution classification (\ref{rq:embed}), \llama is used as a generative model for classification via prompting (\ref{rq:prompt}), and five LMs (three \bert-based models, \xlnet, and \llama) are used as fine-tuned text classifiers specifically tailored to our classification task (\ref{rq:finetune}). 
\rev{This design allows us to systematically examine how language models can be leveraged for the solution-identification task across different modeling paradigms: \textit{embeddings}, \textit{prompting}, and \textit{fine-tuning}. Using LM-generated embeddings as inputs to traditional classifiers allows us to evaluate how well these embeddings capture solution-related content while minimizing additional architectural complexity in the downstream model. This unified setup provides a comprehensive comparison of how these strategies perform in identifying solution-related comments in issue discussions.

Among the LLMs, we selected \llama because it is a widely used open-source model that provides strong performance while remaining accessible for academic experimentation. Compared to other proprietary LLMs, \llama allows full control over inference and fine-tuning pipelines without API restrictions, making it suitable for systematic experimentation and reproducibility.}

We detail each model and its usage to address the research questions (RQs):

\begin{itemize}

    \item \gpt~\cite{achiam2023gpt} is a decoder-only transformer-based model with 1.8T parameters from OpenAI that tokenizes the input using byte pair encoding (BPE) and processes these tokens through multiple layers of self-attention and feed-forward mechanisms. We use \gpt for prompting due to cost constraints: it is substantially cheaper than fine-tuning since. We used another OpenAI model, `text-embedding-3-large', an encoder-style transformer, to generate embeddings that serve as input to the ML models.
    
    \item \llama~\cite{dubey2024llama} is an open-source decoder-only transformer-based model that tokenizes input using SentencePiece BPE. At the time of our experment, \llama were available in two size, 8B paramter and 70B parameter. We used the model with 8B parameters to for finetuning our text-based data to classify comments, and we used the model with 70B parameters to produce embeddings of the comments to be used in ML models and also for prompting to perform the comment classification.
    
    \item \bert~\cite{devlin2018bert} is a bidirectional Transformer model that captures a word’s context from both directions (left-to-right and right-to-left). It tokenizes the input and processes it through layers of self-attention and feed-forward networks. \bert itself is not designed for text generation as it is an encoder-only model. Hence, we use it only for generating embeddings and as a fine-tuned classifier. We used `bert-base-uncased'~\cite{bertbaseun}, which has 110 million parameters.
    
    \item \dbert~\cite{sanh2019distilbert} is a smaller and faster version of \bert, designed to retain 97\% of \bert’s performance while reducing the number of parameters by 40\%. It follows the same tokenization and bidirectional self-attention mechanisms as \bert, generating embeddings from the final hidden states of the tokens. \dbert is not designed for text generation. 
    In our experiments, we used it as a fine-tuned classifier only. We do not use it for generating embeddings to control the number of configurations in our experiments.  We used `distilbert-base-uncased'~\cite{sanh2019distilbert}, which has 67M parameters.
    \looseness=-1
    
    \item \rbert~\cite{liu2019roberta} is an improved version of \bert, with modifications like dynamic masking and longer training on larger datasets. For the same reasons as with \dbert, we use \rbert as a fine-tuned classifier only. We used `roberta-base'~\cite{rbert}, which has 125M parameters.
    
    \item \xlnet~\cite{yang2019xlnet} is an autoregressive model that uses a permutation-based training method, allowing it to capture bidirectional context while maintaining autoregressive properties. It tokenizes the input text and processes it through Transformer layers. Embeddings are generated from the final hidden states, capturing rich contextual information. We use \xlnet as a fine-tuned classifier only. We used `xlnet-base-cased'~\cite{xlnet}, which has 110M parameters.
    
\end{itemize}

All LMs used for classification via fine-tuning were adapted by adding a classification head, which consists of a fully connected layer with a softmax activation function. This layer generates a probability distribution over the two classes: solution and non-solution.

In addition, we use \embtfidf as our embedding baseline method for \ref{rq:embed}: it evaluates how important a word is within an issue report comment compared to a collection of comments (\ie a training set). It calculates the frequency of each comment word (term frequency) and penalizes common words in the training corpus (inverse document frequency). The resulting TF-IDF scores for all words in the comment are combined to create a sparse vector representation, which serves as the final embedding.

\subsection{Model Execution, Training, and Testing}
\label{sub:execution}

\subsubsection{Data Splitting} 
\label{sub:data_split}
\begin{table}[]
\centering
\caption{Dataset Statistics}
\label{tab:data_stats}
\resizebox{.8\columnwidth}{!}{%
\begin{tabular}{c|c|c c c|c}
\toprule
\textbf{Comments} & \textbf{\begin{tabular}[c]{@{}c@{}}Prompt Set\end{tabular}} & \textbf{Training Set} & \textbf{Validation Set} & \textbf{Test Set} & \textbf{Total} \\ 
\midrule
\textbf{\# of Solution}     & 20 & 638  & 71  & 79  & \textbf{808}  \\
\textbf{\# of Non-Solution} & 30 & 3304 & 367 & 408 & \textbf{4109} \\ 
\midrule
\textbf{Total}              & \textbf{50} & \textbf{3942} & \textbf{438} & \textbf{487} & \textbf{4917} \\ 
\bottomrule
\end{tabular}%
}
\end{table}

We selected 50 comments (20 solution and 30 non-solution) from the 4,917 comments for prompt engineering for \llama (see \Cref{sub:experiments}). The remaining 4,867 comments were used for training and testing the MLMs, PLMs, or LLMs through 10-fold cross-validation, where 9 folds were used for training and the remaining fold for testing (see \Cref{tab:data_stats}). Each training and testing set maintained the approximate ratio of solution to non-solution comments found in the full dataset. The same data splits were applied consistently across all models. Hyperparameter tuning was performed for each model, with different strategies used depending on the model type.
\looseness=-1

\subsubsection{Experimental Settings}
\label{sub:experiments}

\textbf{\\ \indent ML Experiments.} Parameter tuning of the ML models was performed on each of the 10 training sets by performing nested 10-fold cross-validation~\cite{cv}. We used cross-validation to ensure that the evaluation of model performance was not dependent on a particular random split of the data. This approach helps mitigate overfitting and provides a more reliable estimate of how the models generalize to unseen data~\cite{Kohavi1995}. Optimal parameters were determined through exhaustive grid search~\cite{gridsearchcv}. Given the imbalanced data, we experimented with \textsc{Smote}~\cite{chawla2002smote}, a technique that generates synthetic examples to balance the minority class (solution comments). \textsc{Smote} was applied only to the training sets. Before generating \embtfidf embeddings, we applied standard text preprocessing, including lowercasing, lemmatization, punctuation removal, and stop word removal. For each of the six ML models (\dtrees, \knns, \lrs, \gnbs, \rfrs, and \svcs), we tested four embedding approaches (\embtfidf, \embbert, \embllama, and \embgpt) and considered both balanced and unbalanced data, resulting in 48 different configurations. The ML models were implemented using the scikit-learn library~\cite{sklearn}. %

\textbf{PLM Experiments.} We applied 10-fold cross-validation to conduct the experiments with each pre-trained language model (PLM). For each fold, we divided the whole data set into training, validation, and test sets and obtained the best hyperparameters using the validation set after fine-tuning the model with the training set. Using the best hyperparameters and the training and validation data sets, we fine-tuned the model and evaluated it with the test set. Fine-tuning followed the recommended \textit{ktrain} guidelines for each model~\cite{Maiya2022Ktrain} 
, focusing on key hyperparameters such as the number of epochs, batch size, and learning rate. To address data imbalance, we experimented with balancing the dataset by duplicating instances of the minority class. This resulted in eight configurations: four models (\bert, \dbert, \rbert, and \xlnet), each trained with both imbalanced and balanced data. The PLM models were implemented using the Ktrain library~\cite{ktrain}.

\textbf{LLM Experiments.} %
We experimented with \llama using \textit{prompting} and \textit{fine-tuning}.

\rev{\textit{Prompting.} For prompt engineering, we used a development set of 50 comments (see \Cref{sub:data_split}), with 43 comments for designing and testing the prompts and 7 used as contextual examples. Prompts were iteratively refined based on their performance in this set. Two authors were involved in the prompt development process. The first author initially designed the prompts by defining the task description, specifying the classification objective, and structuring the prompt to clearly separate the input comment, optional issue description, and the expected output label. These initial prompts were tested on samples from the development set and refined iteratively to improve clarity and model responses. The second author then reviewed the prompts and suggested modifications to the wording, structure, and instructions. Both authors discussed these suggestions and finalized the prompt design, which was subsequently evaluated on the full development dataset.

We explored three widely used prompting strategies in software engineering~\cite{hou2023large}: zero-shot (ZS), few-shot (FS), and chain-of-thought (CoT)~\cite{wei2022chain}. Each resulting prompt had two versions: one with the \textit{issue description} (\ie title and body of the issue) and another without it. Since the issue description discusses the reported problem and its background, it conveys essential contextual information for a model to interpret the discussion comments. Consequently, we investigate to what extent including this information influences classification performance.  

In the few-shot and chain-of-thought prompts, we selected 7 representative examples of solution and non-solution comments from the annotated dataset. To reduce potential selection bias, the examples were chosen to reflect diverse instances of both classes, covering different comment lengths, writing styles, and types of information discussed in the comments. We then experimented with prompts containing 4 and 7 examples to understand the impact of the number of examples on the classification performance. In total, we designed 10 prompts: zero-shot (with and without issue description), few-shot (with 4 and 7 examples, with and without issue description), and chain-of-thought (with 4 and 7 examples, with and without issue description). \Cref{fig:prompt_template} depicts the prompt template for the three prompting strategies.

\begin{figure}[t]
	\centering
	\includegraphics[width=1\linewidth]{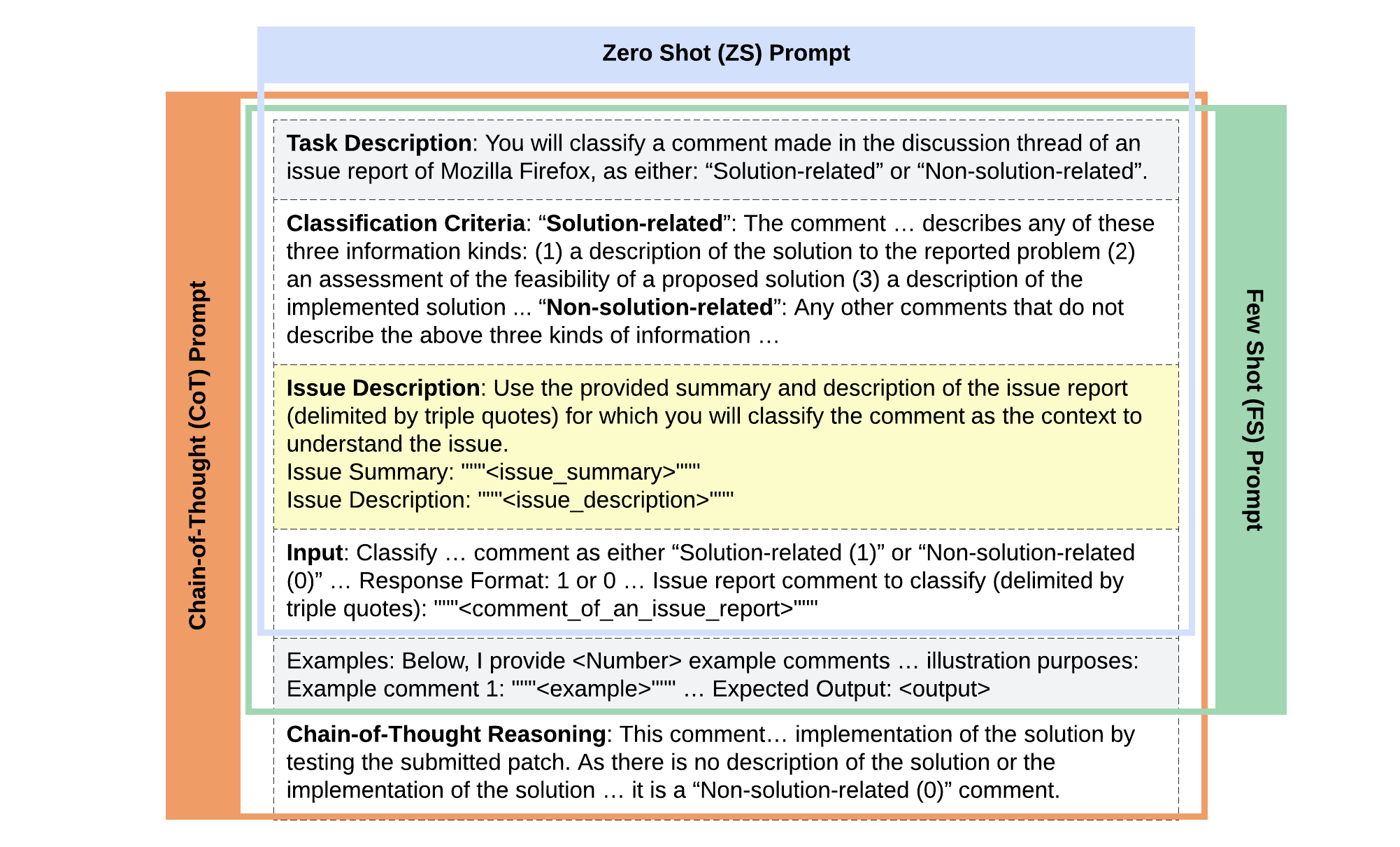}
	\caption{Prompt Templates Based on ZS, FS, \& CoT Prompting (Issue Description is Optional)}
	\label{fig:prompt_template}
\end{figure}

To address non-deterministic \llama responses, we executed the model three times on every comment and averaged the performance of the three executions across the test dataset.}

\rev{\textit{Fine-tuning.}} For fine-tuning, we opted for the \llama-8B model due to resource constraints of our lab, as larger models (\eg with 70B or 405B parameters) demand excessive computational power. However, despite using the smaller 8B version, fine-tuning remained resource-intensive. Hence, we employed Low-Rank Adaptation (LoRA)~\cite{hu2021lora}, a parameter-efficient fine-tuning technique to speed up training. LoRA updates small, low-rank matrices in certain model layers instead of modifying the entire large weight matrices, reducing GPU resource usage. Additionally, we applied 4-bit quantization to decrease model size and computational requirements \cite{dettmers2023qlora}. By reducing the precision of model weights 
to 4-bit, memory usage was significantly lowered.  To reduce computation costs,  we optimized hyperparameters like the number of epochs and batch size on the validation set of the first fold only, then applied the optimal hyperparameters across all other nine folds without tuning.  

\subsubsection{Experiment Execution.}
We executed the experiments in a heterogeneous computing environment, using CPUs for traditional machine learning models (MLMs) and GPUs for pre-trained and large language models (PLMs and LLMs). To reduce training time, especially under cross-validation, we leveraged multiple GPUs in parallel. Specifically, we used three machines: Machine 1 with an AMD EPYC 7532 CPU and an NVIDIA H100 GPU (95 GB), Machine 2 with an AMD EPYC 7543 CPU and eight NVIDIA A40 GPUs (46 GB each), and Machine 3 with an AMD EPYC 9354 CPU and an NVIDIA A100 GPU (41 GB). This setup enabled efficient training and evaluation of models across different architectures and scales.

\subsection{Evaluation Metrics}
\label{sub:metrics}

We evaluated model performance using standard text classification metrics: precision (\textbf{P}), recall (\textbf{R}), and F1 score (\textbf{F1}). Precision measures the proportion of correctly identified solution comments among all those predicted as solutions. It is calculated as: $\text{Precision (\textbf{P})} = \frac{TP}{TP + FP}$, where $TP$ (true positives) are correctly identified solution comments and $FP$ (false positives) are non-solution comments incorrectly labeled as solutions.  Recall assesses a model's ability to detect all actual solution-related comments. It is calculated as: $\text{Recall (\textbf{R})} = \frac{TP}{TP + FN}$, where $FN$ (false negatives) are actual solution comments missed by the model. F1 Score is the harmonic mean of precision and recall, offering a balanced measure between both metrics. F1 Score (\textbf{F1}) is appropriate as we consider both precision and recall equally important: in a practical scenario a model user would want to minimize both false positives and negatives while maximizing true positives. We use the F1 Score to compare and rank the overall performance of the models. If two models have the same F1 score, we prefer the one with higher recall, as it identifies more actual solution comments. Moreover, we defined a comparative metric, \ie\ \textit{Relative Improvement} (\textbf{RI}), to quantify performance gains in precision, recall, or F1 score over a baseline. RI is computed as: $RI = \frac{M_i - M_b}{M_b}$, where $M_i$ denotes the performance metric (precision, recall, or F1 score) of the evaluated model, and $M_b$ represents the corresponding metric of a baseline model.

\section{Results and Discussion}
\label{sec:results}

In this section, we discuss the results and findings of our empirical study on identifying solution comments from developers' discussions. We begin by presenting the results for each research question (RQ), followed by an analysis across different issue report types and problem categories. Finally, we describe the methodology and findings of our qualitative analysis.

\subsection{\ref{rq:embed}: Performance of MLMs with LM Embeddings}
\label{subsec:results_mlms}

\Cref{tab:ml_results_lm_embeddings} shows the classification performance of the six MLMs for solution identification with four LM embeddings and imbalanced \& balanced data. We discuss the results in the following sub-sections.
\looseness=-1

\subsubsection{Effect of Data Balancing} 

\Cref{tab:ml_results_lm_embeddings} shows the performance of MLMs when using different embeddings on both imbalanced and balanced data.  The relative F1 improvement between balanced and imbalanced data (shown in ``DB \textit{vs} no DB'' columns) reveals that some models benefit from data balancing, while others do not. 

\rfr (\rfrs) benefits the most from data balancing, showing substantial F1 improvement (23.1\% to 185\%) across all LM embeddings. These gains stem from a large recall increase (50\% to 282.4\%), accompanied by only a slight precision decrease (7.9\% to 15.3\%), which aligns with expectations for balanced data. As a tree-based model, \rfrs particularly benefits from data balancing since balanced sampling ensures better representation of the minority class during training, leading to improved generalization and recall~\cite{he2009learning}. However, when using \embtfidf embeddings, data balancing causes a 23.6\% F1 drop, primarily driven by a sharp decline in recall (0.633 vs. 0.448).

\knn (\knns) also shows gains from balancing when using \embbert and \embtfidf embeddings, with F1 improvements of 25.5\% and 113.7\%, respectively. On balanced data with \embgpt embeddings, \knns achieves nearly the same F1 performance, with mixed results for precision and recall. However, using \embllama with balanced data results in a significant F1 degradation, mainly due to a sharp precision drop (0.678 vs. 0.316). As a distance-based model, \knns is sensitive to changes in feature space density—balancing methods such as oversampling can distort embedding distributions and neighborhood structures, thereby degrading performance~\cite{stando2024effect,fernandez2018learning}.

Across models, data balancing generally has a negative performance impact compared to no balancing when using \embgpt and \embtfidf, with a few exceptions (\rfrs, \knns, and \dtrees). \embgpt's F1 relative degradation ranges from 0.6\% (\knns) to 9.9\% (\gnbs), while \embtfidf's F1 degradation is higher,  ranging from 9.6\% (\rfrs) to 34.5\% (\gnbs). The effects of \embbert and \llama are more mixed, showing  F1 improvement in some models but deterioration in others. This observation is consistent with prior studies noting that data balancing can change model behavior in unexpected ways, and its effect depends on model architecture and feature representation~\cite{stando2024effect}.

\begin{table}[p]
\centering
\vspace*{4cm}

\rotatebox{90}{
\begin{minipage}{\textheight}
\centering
\caption{Performance of ML Models with Language Model (LM) and \embtfidf (Used as Baseline) Embeddings}
\label{tab:ml_results_lm_embeddings}
\renewcommand{\arraystretch}{1.4}
\setlength{\tabcolsep}{4pt}

\begin{tabular}{cc|ccccc|ccccc|ccccc|cccc}
\toprule
                                                      &                                & \multicolumn{5}{c|}{\textbf{\embbert}}                                                                                                                                                              & \multicolumn{5}{c|}{\textbf{\embgpt}}                                                                                                                                                                                 & \multicolumn{5}{c|}{\textbf{\embllama}}                                                                                                                                                                               & \multicolumn{4}{c}{\textbf{\embtfidf}}                                                                                                            \\
\multirow{-2}{*}{\textbf{Model}}                      & \multirow{-2}{*}{\textbf{DB?}} & \textbf{P} & \textbf{R} & \multicolumn{1}{c|}{\textbf{F1}} & \textbf{\begin{tabular}[c]{@{}c@{}}LM \textit{vs}\\ \embtfidf\end{tabular}} & \textbf{\begin{tabular}[c]{@{}c@{}}DB \textit{vs}\\ No-DB\end{tabular}} & \textbf{P} & \textbf{R} & \multicolumn{1}{c|}{\textbf{F1}}                   & \textbf{\begin{tabular}[c]{@{}c@{}}LM \textit{vs}\\ \embtfidf\end{tabular}} & \textbf{\begin{tabular}[c]{@{}c@{}}DB \textit{vs}\\ No-DB\end{tabular}} & \textbf{P} & \textbf{R} & \multicolumn{1}{c|}{\textbf{F1}}                   & \textbf{\begin{tabular}[c]{@{}c@{}}LM \textit{vs}\\ \embtfidf\end{tabular}} & \textbf{\begin{tabular}[c]{@{}c@{}}DB \textit{vs}\\ No-DB\end{tabular}} & \textbf{P} & \textbf{R} & \multicolumn{1}{c|}{\textbf{F1}}                   & \textbf{\begin{tabular}[c]{@{}c@{}}DB \textit{vs}\\ No-DB\end{tabular}} \\ 
\midrule
                                                      & No                             & 0.305      & 0.748      & \multicolumn{1}{c|}{0.433}       & -15.5\%                                                          &                                                                 & 0.265      & 0.688      & \multicolumn{1}{c|}{0.383}                         & -25.3\%                                                          &                                                                 & 0.327      & 0.564      & \multicolumn{1}{c|}{0.413}                         & -19.2\%                                                          &                                                                 & 0.482      & 0.546      & \multicolumn{1}{c|}{0.512}                         &                                                                 \\
\multirow{-2}{*}{\textbf{\dtrees}}      & Yes                            & 0.326      & 0.484      & \multicolumn{1}{c|}{0.390}       & -24.2\%                                                          & \multirow{-2}{*}{-10.0\%}                                       & 0.331      & 0.433      & \multicolumn{1}{c|}{0.375}                         & -26.9\%                                                          & \multirow{-2}{*}{-1.9\%}                                        & 0.362      & 0.494      & \multicolumn{1}{c|}{0.418}                         & -18.7\%                                                          & \multirow{-2}{*}{1.1\%}                                         & 0.492      & 0.537      & \multicolumn{1}{c|}{\cellcolor[HTML]{FAFFAF}0.514} & \multirow{-2}{*}{0.4\%}                                         \\ 
\midrule
                                                      & No                             & 0.373      & 0.286      & \multicolumn{1}{c|}{0.324}       & 288.8\%                                                          &                                                                 & 0.521      & 0.357      & \multicolumn{1}{c|}{0.424}                         & 409.0\%                                                          &                                                                 & 0.678      & 0.457      & \multicolumn{1}{c|}{\cellcolor[HTML]{FAFFAF}0.546} & 556.1\%                                                          &                                                                 & 0.468      & 0.046      & \multicolumn{1}{c|}{0.083}                         &                                                                 \\
\multirow{-2}{*}{\textbf{\knns}}        & Yes                            & 0.278      & 0.749      & \multicolumn{1}{c|}{0.406}       & 128.3\%                                                          & \multirow{-2}{*}{25.5\%}                                        & 0.279      & 0.858      & \multicolumn{1}{c|}{0.421}                         & 136.7\%                                                          & \multirow{-2}{*}{-0.6\%}                                        & 0.316      & 0.888      & \multicolumn{1}{c|}{0.466}                         & 162.3\%                                                          & \multirow{-2}{*}{-14.6\%}                                       & 0.535      & 0.107      & \multicolumn{1}{c|}{0.178}                         & \multirow{-2}{*}{113.7\%}                                       \\ 
\midrule
                                                      & No                             & 0.425      & 0.699      & \multicolumn{1}{c|}{0.529}       & 5.7\%                                                            &                                                                 & 0.550      & 0.758      & \multicolumn{1}{c|}{\cellcolor[HTML]{FAFFAF}0.638} & 27.4\%                                                           &                                                                 & 0.572      & 0.669      & \multicolumn{1}{c|}{0.617}                         & 23.3\%                                                           &                                                                 & 0.387      & 0.708      & \multicolumn{1}{c|}{0.500}                         &                                                                 \\
\multirow{-2}{*}{\textbf{\lrs}}         & Yes                            & 0.447      & 0.603      & \multicolumn{1}{c|}{0.514}       & 21.0\%                                                           & \multirow{-2}{*}{-2.9\%}                                        & 0.573      & 0.616      & \multicolumn{1}{c|}{0.593}                         & 39.8\%                                                           & \multirow{-2}{*}{-6.9\%}                                        & 0.566      & 0.547      & \multicolumn{1}{c|}{0.557}                         & 31.2\%                                                           & \multirow{-2}{*}{-9.8\%}                                        & 0.443      & 0.407      & \multicolumn{1}{c|}{0.424}                         & \multirow{-2}{*}{-15.2\%}                                       \\ 
\midrule
                                                      & No                             & 0.282      & 0.703      & \multicolumn{1}{c|}{0.402}       & 1.8\%                                                            &                                                                 & 0.332      & 0.797      & \multicolumn{1}{c|}{\cellcolor[HTML]{FAFFAF}0.469} & 18.7\%                                                           &                                                                 & 0.265      & 0.755      & \multicolumn{1}{c|}{0.393}                         & -0.5\%                                                           &                                                                 & 0.289      & 0.623      & \multicolumn{1}{c|}{0.395}                         &                                                                 \\
\multirow{-2}{*}{\textbf{\gnbs}}        & Yes                            & 0.266      & 0.805      & \multicolumn{1}{c|}{0.400}       & 54.7\%                                                           & \multirow{-2}{*}{-0.5\%}                                        & 0.319      & 0.623      & \multicolumn{1}{c|}{0.422}                         & 63.2\%                                                           & \multirow{-2}{*}{-9.9\%}                                        & 0.273      & 0.707      & \multicolumn{1}{c|}{0.394}                         & 52.5\%                                                           & \multirow{-2}{*}{0.4\%}                                         & 0.175      & 0.491      & \multicolumn{1}{c|}{0.259}                         & \multirow{-2}{*}{-34.5\%}                                       \\ 
\midrule
                                                      & No                             & 0.522      & 0.256      & \multicolumn{1}{c|}{0.344}       & -43.3\%                                                          &                                                                 & 0.839      & 0.093      & \multicolumn{1}{c|}{0.167}                         & -72.5\%                                                          &                                                                 & 0.791      & 0.312      & \multicolumn{1}{c|}{0.448}                         & -26.1\%                                                          &                                                                 & 0.581      & 0.633      & \multicolumn{1}{c|}{\cellcolor[HTML]{FAFFAF}0.606} &                                                                 \\
\multirow{-2}{*}{\textbf{\randforests}} & Yes                            & 0.481      & 0.459      & \multicolumn{1}{c|}{0.470}       & -14.2\%                                                          & \multirow{-2}{*}{36.6\%}                                        & 0.730      & 0.354      & \multicolumn{1}{c|}{0.477}                         & -12.9\%                                                          & \multirow{-2}{*}{185.7\%}                                       & 0.670      & 0.468      & \multicolumn{1}{c|}{0.551}                         & 0.6\%                                                            & \multirow{-2}{*}{23.1\%}                                        & 0.705      & 0.448      & \multicolumn{1}{c|}{0.548}                         & \multirow{-2}{*}{-9.6\%}                                        \\ 
\midrule
                                                      & No                             & 0.387      & 0.794      & \multicolumn{1}{c|}{0.521}       & 11.2\%                                                           &                                                                 & 0.629      & 0.701      & \multicolumn{1}{c|}{\cellcolor[HTML]{FAFFAF}0.663} & 41.6\%                                                           &                                                                 & 0.610      & 0.652      & \multicolumn{1}{c|}{0.630}                         & 34.6\%                                                           &                                                                 & 0.356      & 0.684      & \multicolumn{1}{c|}{0.468}                         &                                                                 \\
\multirow{-2}{*}{\textbf{\svcs}}        & Yes                            & 0.408      & 0.772      & \multicolumn{1}{c|}{0.533}       & 49.1\%                                                           & \multirow{-2}{*}{2.4\%}                                         & 0.741      & 0.580      & \multicolumn{1}{c|}{0.651}                         & 81.9\%                                                           & \multirow{-2}{*}{-1.9\%}                                        & 0.672      & 0.579      & \multicolumn{1}{c|}{0.622}                         & 73.8\%                                                           & \multirow{-2}{*}{-1.4\%}                                        & 0.636      & 0.249      & \multicolumn{1}{c|}{0.358}                         & \multirow{-2}{*}{-23.6\%}                                       \\ 
\midrule
                                                      & No                             & 0.382      & 0.581      & \multicolumn{1}{c|}{0.425}       & 41.5\%                                                           &                                                                 & 0.523      & 0.565      & \multicolumn{1}{c|}{0.457}                         & 66.5\%                                                           &                                                                 & 0.540      & 0.568      & \multicolumn{1}{c|}{0.508}                         & 94.7\%                                                           &                                                                 & 0.427      & 0.540      & \multicolumn{1}{c|}{0.427}                         &                                                                 \\
\multirow{-2}{*}{\textbf{\begin{tabular}[c]{@{}c@{}}Avg.\end{tabular}}} 
                                                      & Yes                            & 0.368      & 0.645      & \multicolumn{1}{c|}{0.452}       & 35.8\%                                                           & \multirow{-2}{*}{6.3\%}                                        & 0.495      & 0.577      & \multicolumn{1}{c|}{0.490}                         & 47.0\%                                                           & \multirow{-2}{*}{7.2\%}                                        & 0.477      & 0.614      & \multicolumn{1}{c|}{0.501}                         & 50.3\%                                                           & \multirow{-2}{*}{-1.3\%}                                       & 0.498      & 0.373      & \multicolumn{1}{c|}{0.380}                         & \multirow{-2}{*}{-11.1\%}                                      \\ 
\bottomrule

\multicolumn{21}{c}{
				\textbf{P} = Precision; \textbf{R} = Recall; \textbf{F1} = F1 Score; \textbf{DB} = Data Balancing; \textbf{RI} = Relative Improvement of F1}\\

\end{tabular}

\end{minipage}
}
\end{table}

\subsubsection{Effect of LM Embeddings}

\Cref{tab:ml_results_lm_embeddings}  shows the performance gains achieved by the LM embeddings compared to the baseline \embtfidf method (columns ``LM \textit{vs} \embtfidf''). The table reveals that overall, all the ML models show higher F1 performance when using any LM embeddings, regardless of data balancing, with the exception of \dtrees and \randforests. For these two models, F1 scores degrade consistently across all LM embeddings regardless of data balancing. \rfrs shows a drop ranging from 12.9\% (\embgpt and balancing) to 72.5\% (\embgpt and no balancing) and \dtrees shows drops of 15.5\% (\embbert with no balancing) to 26.9\% (\embgpt with balancing). 
These results suggest that tree-based models struggle to distinguish between solution and non-solution comments when features are encoded as dense semantic vectors, as is typical of LM embeddings. These models perform better on sparse, high-dimensional representations such as \embtfidf, where feature independence allows clearer decision boundaries~\cite{olson2018making}. On the other hand, dense embeddings compress semantic information into fewer dimensions, reducing feature separability for decision trees, which rely on axis-aligned splits rather than geometric similarity~\cite{olson2018making,rana2023comparative}.

Notably, \knns shows the highest F1 improvements when using LM embeddings compared to \embtfidf, with gains ranging from 128.3\% (\embbert with balancing) to 556\% (\embllama with no balancing). This is explained by the low performance of \knns with \embtfidf, which yields F1 scores of 0.083 (without balancing) or 0.178 F1 (with balancing), while LM embeddings produce F1 scores between 0.324 (\embbert without balancing) and 0.546 (\embllama without balancing). \knns's very low performance with \embtfidf stems from the inherent limitations of neighbor‑based classification in high‑dimensional, sparse spaces~\cite{radovanovic2009nearest,prasath2017distance}. In contrast, LM embeddings helped \knns by providing a denser, more meaningful vector space, resulting in large relative improvements compared to \embtfidf.

\begin{figure}[t]
	\centering
	\begin{subfigure}[b]{0.7\textwidth}
		\centering
		\includegraphics[width=\textwidth]{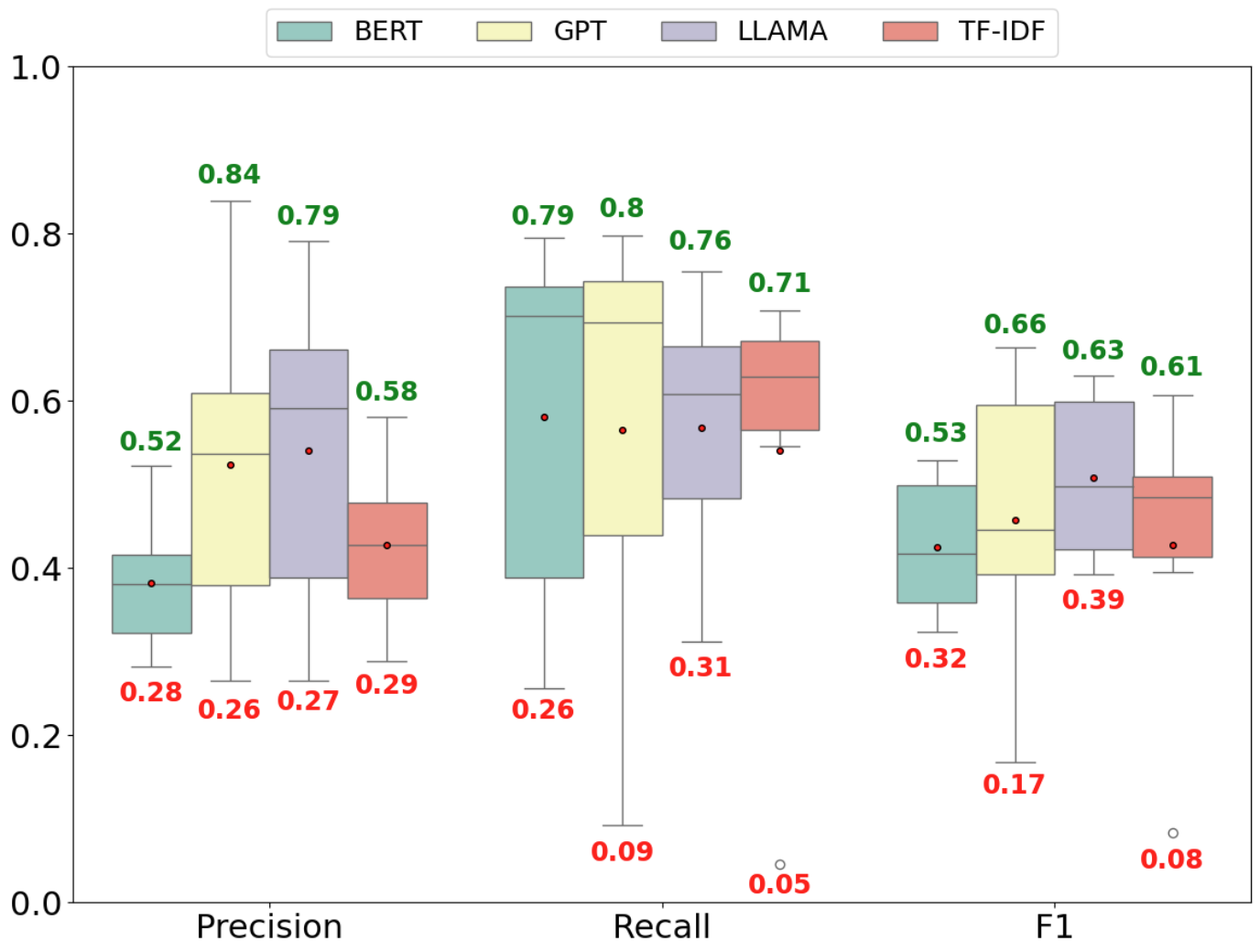}
		\caption{Without Data Balancing (n = 6)}
		\label{fig:lm_embeddings_no}
	\end{subfigure}
	\hfill
	\begin{subfigure}[b]{0.7\textwidth}
		\centering
		\includegraphics[width=\textwidth]{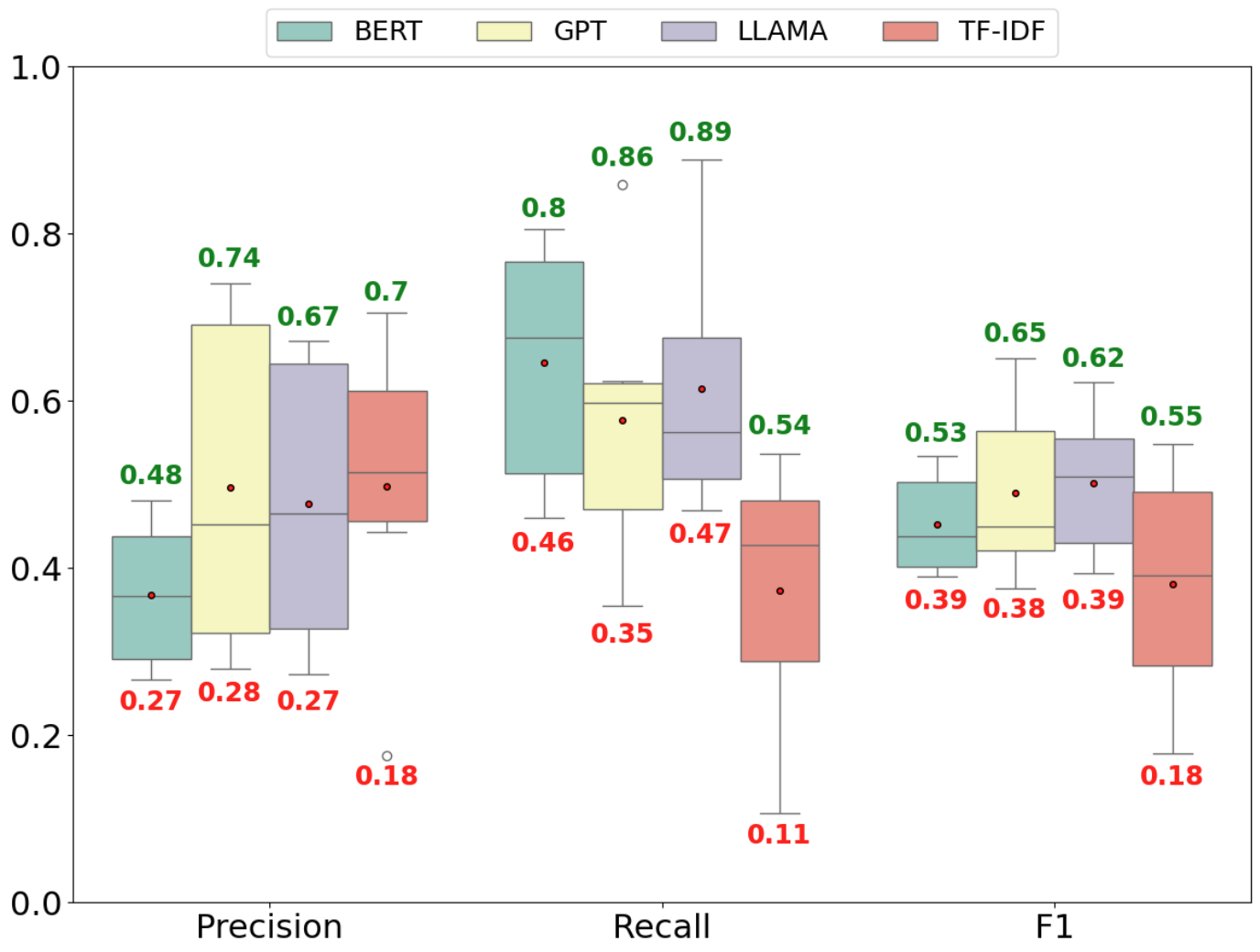}
		\caption{With Data Balancing (n = 6)}
		\label{fig:lm_embeddings_yes}
	\end{subfigure}
	\caption{Performance of the MLMs across LM and \embtfidf Embeddings (Aggregating Across the Six MLMs)
	}
	\label{fig:ml_results}
\end{figure}

\Cref{fig:lm_embeddings_no,fig:lm_embeddings_yes} show MLM performance across the six models when using or not using data balancing, respectively. The figures show that, among the three LM embeddings, \embllama achieves the highest F1 performance on average (0.508 F1 with no balancing and 0.501 with balancing), outperforming \embtfidf's scores (0.427 F1 with no balancing and 0.38 with balancing). This finding becomes more prominent with balancing, where \embllama demonstrates a strong balance between precision and recall across models. In contrast, other LM embeddings and the baseline exhibit less consistency in precision and recall performance across the models.

\subsubsection{Best Performing Configurations} The best-performing configurations of the six MLMs based on F1 score are highlighted in \Cref{tab:ml_results_lm_embeddings} (see the yellow cells). Among the best configurations, \knns, \lrs, \gnbs, and \svcs use LM embeddings, \embgpt or \embllama, while \dtrees and \rfrs use \embtfidf embeddings. Notably, five of the six best MLMs performed best on imbalanced data, with only \dtree requiring balanced data to achieve its highest F1 score.   

The overall best model, \svcs (with \embgpt and no data balancing), achieved the highest F1 score of 0.663, outperforming the second best model by 4.02\% (\lrs with \embgpt and no data balancing). 

\knns saw the most significant F1 score improvement, increasing by 556.1\% using \embllama compared to its \embtfidf configuration, while the best-performing MLM, \svcs, improved its F1 score by 41.6\% using \embgpt over \embtfidf.

\rqanswer{\textbf{\ref{rq:embed} Findings}: Different ML models exhibit varying levels of effectiveness depending on the type of embeddings and the use of data balancing. Among the embeddings, \embllama produces the most consistent improvements, yielding the highest average F1 across models and balancing precision–recall trade-offs. While data balancing helps certain models—particularly \rfrs and \knns—it has limited or even adverse effects on others, depending on how each model processes dense embedding representations. Tree-based models (\dtrees, \rfrs) tend to perform better with sparse features (\ie \embtfidf), whereas distance- and margin-based models (\knns, \svcs) benefit more from dense LM-based embeddings. Overall, these findings indicate that LM embeddings can improve traditional classifiers, but their effectiveness ultimately depends on how well the model architecture aligns with the representational properties of the embeddings.\looseness=-1}

\subsection{\ref{rq:prompt}: Performance of Prompted Language Models}

\subsubsection{Impact of Prompting Strategies}

\Cref{tab:prompt_results} presents \llama's performance with the 10 prompts we designed based on different prompting strategies. The best three prompts, identified by the F1 score (highlighted in yellow in the table), include the best prompt for each strategy: zero-shot (ZS), few-shot (FS), and chain-of-thought (CoT). These prompts show comparable performance,  with CoT performing the best, achieving an F1 score of 0.517 when using no context and seven examples.

None of the top three prompts incorporates issue description. Including the issue description reduces the F1 score by 6\% (ZS), 6.8\% (FS), and 8.5\% (CoT),  primarily due to significant precision degradation (8.7\% to 12.4\%) that outweighs the recall improvement (6.2\% to 7.6\%). For CoT, adding context results in lower precision (by 8.7\%) and recall (by 7.9\%).

In FS and CoT, prompts with seven examples outperformed those with four examples by 2.2\% (FS) and 1.8\% (CoT) in F1 score. Notably, precision increased by 4.3\% (FS) and 6.3\% (CoT) with seven examples, though recall decreased by 1.7\% (FS) and 5.6\% (CoT). Overall, using more examples in the prompt helped \llama better identify solution-related content.
\looseness=-1

\llama's relatively low performance via prompting (compared to the best MLMs presented in \Cref{subsec:results_mlms}) highlighted the challenge of identifying solution-related content by generative models. While FS and CoT have outperformed ZS in various generation-related SE tasks~\cite{hou2023large}, this trend does not hold for identifying solution content, where all strategies yield similar results. We believe this difficulty arises from the diverse ways in which solutions can be described, as observed during issue annotation. Although FS and CoT include examples with or without reasoning, they do little to capture the various patterns of solution discussions in issue report comments. To effectively classify solution comments, a model must learn these diverse patterns, which explains why MLMs (see \Cref{subsec:results_mlms}), PLMs, and LLMs (see \Cref{subsec:results_lms}) trained or fine-tuned specifically for this task achieve better results in identifying solution content. This finding aligns with prior software engineering studies showing that fine-tuning generally yields superior performance to prompting for tasks such as code generation~\cite{weyssow2025exploring,shin2023prompt}, code summarization~\cite{ahmed2022few}, and automated code review~\cite{pornprasit2024fine}.

\begin{table}[t]
	\centering
	\caption{Performance of \llama with Prompting}
	\label{tab:prompt_results}
	\resizebox{0.85\columnwidth}{!}{%
		\begin{tabular}{ccc|ccc}
			\toprule
			\textbf{\begin{tabular}[c]{@{}c@{}}Prompting\\ Strategy\end{tabular}} &
			\textbf{\begin{tabular}[c]{@{}c@{}}Issue\\ Description\end{tabular}} &
			\textbf{\begin{tabular}[c]{@{}c@{}}\# of\\ Examples\end{tabular}} &
			\textbf{P} & \textbf{R} & \textbf{F1} \\
			\midrule
			\multirow{2}{*}{\textbf{Zero Shot (ZS)}} & No  & - & 0.406 & 0.700 & \cellcolor[HTML]{FAFFAF}0.514 \\
			& Yes & - & 0.356 & 0.753 & 0.483 \\
			\midrule
			\multirow{4}{*}{\textbf{Few Shot (FS)}} & \multirow{2}{*}{No}  & 4 & 0.377 & 0.742 & 0.500 \\
			&                      & 7 & 0.394 & 0.729 & \cellcolor[HTML]{FAFFAF}0.511 \\
			\cmidrule(lr){2-6}
			& \multirow{2}{*}{Yes} & 4 & 0.341 & 0.779 & 0.474 \\
			&                      & 7 & 0.344 & 0.774 & 0.476 \\
			\midrule
			\multirow{4}{*}{\textbf{Chain of Thought (CoT)}} & \multirow{2}{*}{No}  & 4 & 0.390 & 0.727 & 0.507 \\
			&                      & 7 & 0.414 & 0.686 & \cellcolor[HTML]{FAFFAF}0.517 \\
			\cmidrule(lr){2-6}
			& \multirow{2}{*}{Yes} & 4 & 0.324 & 0.800 & 0.462 \\
			&                      & 7 & 0.378 & 0.632 & 0.473 \\
			\bottomrule
		\end{tabular}%
	}
\end{table}

\subsubsection{Impact of Issue Description and Non-Deterministic Responses}

Using issue description in the prompts proved ineffective, as it lowered precision by an average of 12.01\% while increasing recall only by 4.29\%, leading to an overall decrease of 7.11\% in F1 score (across all designed prompts). 

We hypothesize that including the issue description as context overloaded \llama with too much information (\eg task description, context, target comment, output format, and examples with labels). This hypothesis is supported by a deeper analysis of \llama's generated responses. Despite designing the prompts for binary output (0 or 1), \llama generated non-binary responses in 2,674 cases out of a total of 146,010 responses (for the three \llama executions). Notably, 2,662 of these non-binary responses (99\%) came from prompts that included context.

We also examined the non-deterministic LLM behavior by running each prompt three times. \llama produced consistent classification responses, with only 950 out of 48,670 sets (1.95\%) showing any variation. This minor inconsistency had a negligible effect on overall performance, as the average standard deviation of F1 scores across all three executions was just 0.0049.

\rqanswer{\textbf{\ref{rq:prompt} Findings}: \llama\unskip’s overall low performance with prompting highlights the challenge generative models face in identifying solution-related content. Among the experimented strategies, CoT with seven examples performs the best (F1 = 0.517), yet all prompting approaches struggle to capture the diverse ways solutions are expressed in issue discussions. Including issue descriptions further reduces precision, suggesting that excessive contextual information may hinder rather than help model understanding. Overall, these findings underscore the limited effectiveness of prompting for the solution identification task.}

\subsection{\ref{rq:finetune}: Performance of Fine-Tuned Language Models}
\label{subsec:results_lms}

\subsubsection{Pre-trained Language Models (PLMs)}

\Cref{tab:ft_results} shows the performance of fine-tuned PLM models on both balanced and imbalanced data. Among the four models, \rbert achieved the highest F1 scores for both imbalanced (0.689) and balanced (0.692) data, with a maximum relative improvement of 5.8\% over \dbert with data balancing (the 2nd best model). This performance is driven by \rbert's high recall (0.704 without balancing and 0.674 with balancing). 

While \bert had the highest precision (0.690), \xlnet achieved the highest recall (0.713), both with balanced data. These results highlight how different models excel under different configurations.

Data balancing had minimal impact on the three \bert-based models, with only small variations in precision and recall. However, \xlnet showed a more prominent effect with data balancing: its precision dropped by 5.4\%, while recall improved by 6.4\%. On balanced data, the \bert-based models outperformed \xlnet in precision (by 9.2\% to 9.4\%).  In contrast, \xlnet led in the recall, surpassing \rbert by 0.18\%, \bert by 8.7\%, and \dbert by 11.9\%. On imbalanced data, \xlnet had the lowest precision (0.651) but the second-highest recall (0.670), demonstrating different behavior compared to the \bert-based models.

\begin{table}[t]
	\centering
	\caption{Performance of LMs with Fine-Tuning}
	\label{tab:ft_results}
	\resizebox{0.5\columnwidth}{!}{%
		\begin{tabular}{cc|ccc}
			\toprule
			\textbf{LMs} & \textbf{DB} & \textbf{P} & \textbf{R} & \textbf{F1} \\
			\midrule
			\multirow{2}{*}{\textbf{\bert}}   & No  & 0.690 & 0.649 & 0.668 \\
			& Yes & 0.672 & 0.656 & 0.664 \\
			\midrule
			\multirow{2}{*}{\textbf{\dbert}}  & No  & 0.681 & 0.641 & 0.660 \\
			& Yes & 0.672 & 0.637 & 0.654 \\
			\midrule
			\multirow{2}{*}{\textbf{\rbert}}  & No  & 0.675 & 0.704 & 0.689 \\
			& Yes & 0.674 & 0.712 & 0.692 \\
			\midrule
			\multirow{2}{*}{\textbf{\xlnet}}  & No  & 0.651 & 0.670 & 0.660 \\
			& Yes & 0.616 & 0.713 & 0.661 \\
			\midrule
			\textbf{\llamaft} & Yes & 0.626 & 0.838 & 0.716 \\
			\bottomrule
		\end{tabular}%
	}
\end{table}

\subsubsection{Large Language Models (LLMs)}
Given the significant time required to fine-tune \llama for one configuration (approximately three days on our machines), we adopted a practical approach for cross-validation. We performed hyperparameter tuning using one fold to identify the best parameters: a batch size of 8, 5 epochs, and the use of balanced data. Using this configuration, we fine-tuned \llama on the training set of all 10 folds and measured its performance on the corresponding test sets.

As shown in \Cref{tab:ft_results}, \llamaft achieved the highest F1 score of 0.716, outperforming all PLMs and MLMs (see \Cref{tab:ml_results_lm_embeddings}). It also yielded the highest recall of 0.838 and a precision of 0.626. Compared to the best PLM, \rbert, on the balanced data, \llamaft improved recall by 17.7\% and lowered precision by 7.1\%, resulting in a 3.5\% F1 increase. This highlights \llama's superior performance in identifying solution-related content. \llamaft generated more true positives (660 vs. 561) and fewer false negatives (128 vs. 227) compared \rbert, illustrating its better ability to learn relevant data patterns for the classification of solution content.

\rqanswer{\textbf{\ref{rq:finetune} Findings}: Among the fine-tuned models, \rbert achieves the best overall performance among the PLMs, while \llamaft surpasses all models in identifying solution-related content. Data balancing has minimal effect on the \bert-based models but leads to a notable recall–precision tradeoff for \xlnet. Compared to the best PLM (\ie\ \rbert), \llamaft attains substantially higher recall with only a slight precision drop, resulting in a clear F1 gain. These findings indicate that large language models, even with limited fine-tuning, better capture the linguistic patterns associated with solution-related content than smaller pre-trained models.}

\subsection{\ref{rq:ensemble}: Performance of Ensemble Models}

Given that the classifiers we evaluated use various approaches and algorithms to learn patterns in the data, we conducted additional experiments to assess their combined predictive capabilities.

We implemented a straightforward majority voting approach, where the class of a comment is determined by the number of classifiers that predict it as either a solution or a non-solution. The class with the most votes is selected. In the event of a tie, we assessed two strategies: a conservative approach that predicts non-solution, due to insufficient evidence for a solution prediction, and a non-conservative approach that predicts a solution, aiming to enhance recall by increasing the likelihood of identifying true solution comments.

We designed four categories of ensemble models based on the six best MLMs, the four fine-tuned PLMs, and the two best-performing LLMs (fine-tuned and prompted \llama), applying both conservative and non-conservative strategies as appropriate.

\begin{itemize}
    \item \textbf{Model Type:} In this category, we combined models within and across the three model types. Specifically, we built three ensemble models using only MLMs, PLMs, and LLMs, and four additional ensemble models by combining different model types—\ie MLMs + PLMs, MLMs + LLMs, PLMs + LLMs, and MLMs + PLMs + LLMs. Since each ensemble model includes an even number of models, both conservative and non-conservative versions were designed, resulting in 14 ensemble models (\eg \textit{PLMs-Conservative}, \textit{MLMs+LLMs-Non-Conservative}).

    \item \textbf{Top K Performers:} To investigate whether top-performing individual models collectively yield better performance, we sorted the 12 best-performing models by their F1 scores and incrementally grouped them from the top of the list. Starting with the top two, each subsequent ensemble model added the next best model. Ensemble models with an even number of models included both conservative and non-conservative versions, producing 17 ensemble models in total (\eg\ \textit{Top-12-Conservative}, \textit{Top-12-Non-Conservative}, \textit{Top-7}).  

    \item \textbf{Bottom K Performers:} Complementing the previous category, we also designed ensemble models using the lowest-performing classifiers from the bottom of the ranked list. Starting with the bottom two models, each subsequent ensemble added the next higher-ranked model. The goal was to explore whether combining weaker models could still enhance performance and to compare against the ensemble models of the \textit{Top K Performer} category. This category includes 15 ensemble models, \eg \textit{Bottom-11}, \textit{Bottom-10-Conservative}, \textit{Bottom-10-Non-Conservative}.  

    \item \textbf{Best of Each Model Type:} Finally, to assess whether a small yet diverse ensemble can perform competitively, we combined the best model from each type: \svcs (MLM), \rbert (PLM), and \llamaft (LLM).  
\end{itemize}

\begin{table}[t]
\centering
\caption{Performance of Ensembled Models}
\label{tab:ensembled_results}
\resizebox{1\columnwidth}{!}{%
\begin{tabular}{cc|cc|cc|cc}
\toprule
\textbf{Combination Category}                 & \textbf{Best Combinations} & \textbf{P} & \textbf{RI} & \textbf{R} & \textbf{RI} & \textbf{F1} & \textbf{RI} \\ 
\midrule
\textbf{Baseline}                             & \llamaft                   & 0.626      & -           & 0.838      & -           & 0.716       & -           \\ 
\midrule
\multirow{2}{*}{\textbf{Model Type}}          & MLMs + PLMs                & 0.689      & 10.19\%     & 0.732      & -12.6\%     & 0.710       & -0.8\%      \\
                                              & PLMs + LLMs                & 0.661      & 5.64\%      & 0.774      & -7.6\%      & 0.713       & -0.4\%      \\ 
\midrule
\multirow{2}{*}{\textbf{Top K Performers}}    & Top-7                      & 0.713      & 13.90\%     & 0.742      & -11.4\%     & 0.727       & 1.5\%       \\
                                              & Top-5                      & 0.709      & 13.37\%     & 0.740      & -11.7\%     & 0.724       & 1.1\%       \\ 
\midrule
\multirow{2}{*}{\textbf{Bottom K Performers}} & Bottom-11                  & 0.710      & 13.41\%     & 0.701      & -16.4\%     & 0.705       & -1.6\%      \\
                                              & Bottom-10                  & 0.668      & 6.75\%      & 0.730      & -12.9\%     & 0.697       & -2.6\%      \\ 
\midrule
\textbf{Best of Each Model Type}              & \svcs + \rbert + \llamaft  & 0.702      & 12.26\%     & 0.775      & -7.4\%      & 0.737       & 2.9\%       \\ 
\bottomrule

\multicolumn{8}{c}{
    \textbf{RI} = Relative Improvement compared to the baseline (\llamaft)
}\\
\end{tabular}%
}
\end{table}

\Cref{tab:ensembled_results} shows the performance of the top two ensemble models from each category, out of 47 ensemble models. The results of all models are found in our replication package~\cite{repl_pack}. We used the best individual model as the baseline: \llamaft.
\looseness=-1

None of the 14 combinations by  \textit{model type} surpassed the baseline in terms of F1 score. Although 13 combinations improved precision, their recall dropped, resulting in no overall F1 improvement. The best combinations in this category were PLMs with MLMs (F1: 0.710) and PLMs with LLMs (F1: 0.713), both showing precision improvements of 10.2\% and 5.6\%, respectively, but failing to surpass the baseline F1 score of 0.716 due to a decrease in recall.

As for \textit{top K performers}, 16 of 17 combinations enhanced precision, while one improved recall, but none improved both metrics simultaneously. Despite this, 11 out of 17 combinations slightly outperformed the baseline, with F1 improvements of 0.16\% to 1.5\%. The best results were achieved using the top 7 or top 5 models, which exceeded the baseline by 1.5\% and 1.1\% F1, respectively. 

Ten of 15 \textit{bottom K performer} combinations boosted precision, but only one improved recall, resulting in no overall F1 improvement over the baseline. The best configurations in this category involved the bottom 11 or 10 models.

The highest-performing combination was obtained by selecting \textit{the best model from each type}: \svc from MLMs, \rbert from PLMs, and \llamaft from LLMs. This combination highlights the value of using diverse models to identify solution-related content. It improved precision by 12.3\%, while recall decreased by 7.4\% (the smallest recall drop among the 12 combinations that outperformed the baseline), leading to a 2.9\% F1 improvement. The slight performance gain, even after combining the best models from each type, underscores the inherent difficulty of this task and indicates that further research is needed to develop more effective ensembling or modeling strategies.

\rqanswer{\textbf{\ref{rq:ensemble} Findings}: Combining different models leads to modest performance improvements, with the combination of the best models from each type (\svc, \rbert, and \llamaft) achieving the highest F1 gain of 2.9\% compared to \llamaft. Most other ensemble configurations either fail to surpass \llamaft or improve only precision at the cost of recall. These results indicate that identifying solution-related content remains a challenging task, even when leveraging complementary model architectures. The limited gains suggest that further research is needed to explore more effective ensembling techniques or novel modeling approaches to better capture the complex patterns in solution comments.}

\subsection{\ref{rq:types}: Results Across Issue  Report Types and Problem Categories}
\label{sec:results_across_prob_cat}

We assessed the performance of the 12 best individual models across different issue report types (defects, enhancements, tasks) and problem categories (UI issues, crashes, code improvements, \etc).

\textbf{Issue Report Types.}
Out of the 12 models, eight achieved the highest F1 score for \textit{tasks}, three for \textit{defects}, and one for \textit{enhancements}. The best-performing model, \llamaft, achieved its highest F1 score for tasks (0.758) compared to defects and enhancements (0.715 each). On average, tasks had a 6.4\% (median 10.5\%) and 4.6\% (median 5.8\%) F1 improvement over defects and enhancements.

Tasks also had higher average precision (0.605 vs. 0.580 and 0.594) and recall (0.710 vs. 0.676 and 0.678) than defects and enhancements. This trend was consistent across all model types (MLMs, PLMs, and LLMs), suggesting that solution-related content is easier to identify in tasks, while it is most challenging in the other issue types.

\textbf{Problem Categories.}
During the annotation process in our prior work~\cite{saha2025decoding}, we classified the issues into 17 problem categories, such as UI issues, crashes, and code improvements (see \Cref{tab:prob_categories}). The best-performing model, \llamaft, struggled to identify solution-related content in certain categories. For example, its F1 score was lowest for "incorrect page rendering" (0.596) and "unnecessary code removal" (0.615), compared to higher-performing categories like "audit tasks" (0.923), "error handling improvement" (0.8), and "code improvement" (0.783). This trend was consistent for both precision and recall.

To further investigate this trend, we analyzed the combined results of all 12 models and their different types (MLMs, PLMs, LLMs). Categories like "incorrect page rendering," "unnecessary code removal," and "preventive changes" consistently ranked lower in average F1 scores, while "audit tasks," "error handling improvement," "code improvement," and "compatibility issues" ranked higher. For instance, across all models, "error handling improvement" had the highest average rank (3.8), while "incorrect page rendering" had the lowest (15.8).

\rqanswer{\textbf{\ref{rq:types} Findings}: Despite architectural differences, the 12 models exhibit similar behavior when identifying solution-related content across issue types and problem categories. The observed performance differences likely stem from the higher linguistic variability used to describe solution and non-solution content in defects, enhancements, and certain problem categories (\eg incorrect page rendering and unnecessary code removal). This variability makes it more difficult for the models to learn consistent patterns from the data for accurate classification.}

\subsection{Qualitative Analysis}
\label{sec:qualitative_analysis}

To gain deeper insights into the models’ decision-making behavior, we qualitatively analyzed the predictions by the three best-performing models of each category: \llamaft, \rbert, and \svcs. The goal of this analysis was to understand how these models interpret and classify comments as \textit{solution} or \textit{non-solution} in challenging cases. Specifically, we focused on two scenarios: (1) when all models failed to correctly predict a comment, and (2) when the best model (\ie\ \llamaft) failed, but the other two models (\ie\ \rbert and \svcs) succeeded. Through this analysis, we aimed to uncover linguistic and contextual patterns that may influence the models' predictions and to characterize the nature of their misclassifications.

We formulated two research questions (RQs) to guide our qualitative analysis:
\begin{enumerate}[label=\textbf{RQ$_\arabic*$:}, ref=\textbf{RQ$_\arabic*$}, itemindent=0cm,leftmargin=1cm, start=6]
	\item \label{rq:all-fail} Why do the best-performing models fail to predict a comment as (non)-solution?
	\item \label{rq:llama-fail} Why does \llamaft fail to predict a comment as (non)-solution while \rbert and \svcs succeed?
\end{enumerate}

\subsubsection{Methodology}
\label{sub:qual_method}

We qualitatively examined the \textit{false positive (FP)} and \textit{false negative (FN)} cases produced by the three best-performing models: \llamaft, \rbert, and \svcs. Among the 4,867 total comments, all three models failed to predict the correct label for 183 comments (108 FPs and 75 FNs). In addition, \llamaft failed to predict correctly for 193 comments (171 FPs and 22 FNs) where both \rbert and \svcs succeeded. To ensure the analysis was both representative and manageable, we applied stratified random sampling with a 90\% confidence level and a 10\% margin of error on the 183 and 193 comments. This yielded 50 comments (32 FPs and 18 FNs) for \ref{rq:all-fail}, and 51 comments (43 FPs and 8 FNs) for \ref{rq:llama-fail}. These sampled comments formed the basis of our manual evaluation.

We adopted an \textit{open coding} approach~\cite{spencer2009card} to analyze the failure cases. Two researchers independently reviewed each sampled comment to determine potential reasons behind the models’ incorrect predictions. When necessary, they also examined the broader discussion within the issue report to account for the full context and information expressed in the discussion. During the analysis, the researchers documented the potential reasons for misclassification along with detailed notes in a shared spreadsheet. The analysis focused on several comment-level characteristics that might influence model behavior, such as linguistic structure, semantic ambiguity, use of domain-specific terminology, dependency on additional context (\eg issue description or surrounding comments), and the presence or absence of solution-related clues. After the independent coding phase, the researchers met to discuss and reconcile their findings, arriving at a consensus for each case. Similar reasons were grouped into broader categories representing distinct causes of model failure. Inter-rater agreement between the two researchers was strong, with an agreement rate of 89.1\% (Cohen’s Kappa~\cite{Cohen}, $\kappa = 0.89$, indicating almost perfect agreement~\cite{viera2005understanding}; Krippendorff’s Alpha~\cite{krippendorff2018content}, $\alpha = 0.95$, indicating highly reliable coding consistency~\cite{marzi2024k}).

\subsubsection{\ref{rq:all-fail} Findings}
\label{sub:result_all_fail}

By analyzing the comments where all three models failed to classify correctly, we found several recurring patterns that illustrate the challenges of distinguishing solution and non-solution comments by the models.

\textbf{Non-Solution Comments Predicted as Solution (FPs = 32).}
We identified four major patterns that led the models to classify non-solution comments as solutions incorrectly. The most frequent pattern involved comments that were related to the \textit{implemented solution} but did not actually propose, evaluate, or describe a solution themselves (9 cases). Although these comments contained technical terms or code-related expressions resembling solution statements, their purpose was not to suggest or explain a new fix. Instead, they discussed aspects of an already implemented solution, which likely confused the models. Specifically, these comments included: (i) reviews or suggestions on the implemented solution (4 cases), (ii) verification statements confirming that the implemented solution works as expected (2 cases), (iii) explanations of how the implemented solution functions (2 cases), and (iv) specific requirements for implementing a solution (1 case). For example, in the comment \textit{``We could avoid spreading the use of the horribly-named 'pageproxystate' attribute and use some other name''}, the developer is providing feedback on the implemented solution rather than describing or proposing a solution for the respective issue.

Another frequent pattern was a \textit{lack of sufficient context} for predicting a comment as non-solution (8 cases). For example, a comment may describe a solution for a collateral problem (\ie not directly related to the target issue) or refer to a fix for a different but related issue. To correctly interpret such cases, the models would need access to the surrounding discussion and/or issue title and description. However, since each comment was classified in isolation, the lack of contextual knowledge led to misclassification.

The third most common pattern was the presence of certain \textit{keywords} (\eg ``created attachment'', ``solution'', ``implement''), which appeared in six comments. The models seemed to overemphasize these words, treating them as strong indicators of a solution even when they were used in unrelated contexts. For example, a developer might attach an image or log file to illustrate the problem, leading the model to misclassify the comment as a solution simply because it contains the phrase ``created attachment''. This keyword likely misleads the models because actual solution comments that include implemented patches also contain similar phrases to refer to the patch, causing the model to associate the term strongly with solution-related content. 

Finally, we observed three cases that were \textit{hard even for humans} to decide. These comments tended to be lengthy, ambiguous, and filled with technical details, which likely prompted the models to misinterpret them as solution-related. We did not observe any specific patterns for the remaining six comments.

\textbf{Solution Comments Predicted as Non-Solution (FNs = 18)}.  
For the missed solutions, we found that the majority of errors (\textit{12 of 18}) occurred due to a \textit{lack of contextual information}. These comments were typically short, abstract, or discussed the solution at a high level. To understand these comments, the models require contextual information, such as the issue description and surrounding comments. Since the models classify each comment independently, they failed to recognize the implicit solution clues (\eg keywords, technical terms, \etc) present in these cases. We also observed that when comments contained little to no developer-written text, models tended to label them as non-solutions even when they actually described solutions (3 cases). These comments typically included \textit{auto-generated content} (\eg automated commit or merge messages, patch links, \etc) that referenced the implemented solution.
Lastly, the models often missed \textit{implicit solutions} expressed as questions (2 cases), \ie when a developer asks if their idea could solve the problem. Such comments convey a potential fix but lack explicit phrasing, which makes them difficult for the models to identify as solutions. For the remaining comment, we did not observe any pattern.

\rqanswer{\textbf{\ref{rq:all-fail} Findings}: The best models from each type (\ie\ \svc, \rbert, and \llamaft) struggle to distinguish solution from non-solution comments when they contain misleading clues (\eg keywords or code snippets) or lack contextual information (\eg issue description and surrounding discussion) needed to interpret the comment accurately. Non-solution comments with technical jargon or references to implementation details often lead to false positives, while concise or implicitly phrased solution comments are frequently missed. These findings suggest that models need to incorporate contextual information from the issue description and surrounding comments to identify solution-related content more accurately.\looseness=-1}

\subsubsection{\ref{rq:llama-fail} Findings} 
By examining the comments where only \llamaft failed while the other two models succeeded, we identified several patterns that illustrate the weaknesses of the LLM-based classifiers compared to MLMs and PLMs.

\textbf{Non-Solution Comments Predicted as Solution (FP = 43).}
The majority of \llamaft\unskip’s false positives stemmed from its strong dependence on technical language rather than contextual understanding (26 cases). The model seemed to rely heavily on programming-related texts (\eg class/method/variable names) (13 cases), technical artifacts (\eg patch links, stack traces) (11 cases), specific keywords (\eg ``patch'', ``fixes this problem'', ``created attachment'') (7 cases), and source code snippets (3 cases). Note that one comment can include multiple elements. The presence of such technical elements appeared to mislead the model into assuming the presence of a solution, even when the comment merely described supporting information or debugging outputs. For example, the comment \textit{``<source\_code>. So the site assumes that if .csstransforms3d is set then the browser must be either WebKit or implement unprefixed 3D transforms...''} includes multiple technical elements—such as a code snippet, programming-related text (\ie \textit{.csstransforms3d'}), and the keyword \textit{implement'}. However, rather than proposing a solution, the comment actually explains the underlying cause of the problem, which likely led \llamaft to misclassify it as a solution.

Another pattern for false positives of \llamaft was the model's tendency to misinterpret comments related to the \textit{implemented solution} as actual solutions. For instance, it frequently classified \textit{code review} comments (6 cases) or \textit{solution verification} comments (3 cases) as solution descriptions, likely because they contained language resembling implementation activities. In these cases, the model failed to discern that the comment was reviewing or verifying an existing code change rather than proposing, assessing, or describing a solution as discussed in \Cref{sub:result_all_fail}. We also found that \llamaft misclassified comments due to a \textit{lack of sufficient context} (8 cases). Comments describing collateral problem analyses, implementation requirements, or general work updates were mistakenly treated as solution-related, likely because the model lacked the broader discussion context needed to interpret their intent. In contrast, the MLM and PLM correctly identified these as non-solution comments, suggesting that their supervised training and fine-tuning helped them better learn the boundaries of what constitutes a ``solution'' within this domain.

\textbf{Solution Comments Predicted as Non-Solution (FN = 8).}
\llamaft struggled with brief or abstract descriptions of solutions. Most errors (6 of 8) resulted from comments that lacked sufficient textual detail—such as high-level summaries (\eg commit messages or short implementation notes) or auto-generated messages referencing fixes. For example, the comment \textit{``This is a regression from bug 1043644 part 1, which switched to using the content viewer bounds instead of the frame bounds to fix behaviour on metro. Switching back to the frame bounds fixes the problem.''} concisely explains what was changed to resolve the issue, but does not elaborate on the implementation details. Such high-level phrasing can make it difficult for the model to distinguish these as solutions. The model also failed when a comment contained both problem and solution descriptions, often focusing on the problem portion and overlooking the solution intent. Meanwhile, the other models succeeded, likely because their training exposed them to similar examples, enabling them to capture subtle solution-related patterns such as programming-related terms or ``fix''-related phrasing.

\rqanswer{\textbf{\ref{rq:llama-fail} Findings}: \llamaft struggles to distinguish solution comments when explicit linguistic or contextual information is limited. Its strong reliance on technical language (\eg class/method/variable names, keywords, code snippets) causes it to misclassify non-solution comments containing code or debugging information, while concise or high-level solution comments are often missed. These findings suggest that \llamaft benefits from more in-domain labeled examples to effectively capture the subtle patterns of solution-related comments. In contrast, smaller MLM and PLM models, fine-tuned on supervised data, perform better in these cases, likely because their training allows them to learn domain-specific linguistic boundaries more efficiently than the larger, more generalized LLM.}

\section{Generalizability Study}
\label{sec:generalizability}

To assess the generalizability of the models trained/fine-tuned on Mozilla issue reports, we evaluated their performance on issue reports from two other projects: Chromium~\cite{chromium} and GnuCash~\cite{gnucash}. The datasets for these projects were obtained from our prior study~\cite{saha2025decoding} and used to test the best-performing models from each category: \svcs (MLMs), \rbert (PLMs), and \llamaft (LLMs), under different training/fine-tuning settings. This generalizability study is designed to address the following research question (RQ):

\begin{enumerate}[label=\textbf{RQ$_\arabic*$:}, ref=\textbf{RQ$_\arabic*$}, itemindent=0cm,leftmargin=1cm, start=8]
	\item \label{rq:gener}{How well do models fine-tuned on Mozilla data generalize to other projects?}
\end{enumerate}

\subsection{Dataset}

\rev{In our prior work~\cite{saha2025decoding}, we compared the issue resolution process performed at Mozilla Firefox with that employed at Chromium~\cite{chromium} and GnuCash~\cite{gnucash} through a qualitative analysis of additional issue reports. In that work, these projects were selected because they represent widely used open-source systems with active issue-tracking practices and publicly accessible discussions and we aimed to compare Firefox with one system of the same domain and one system of a different domain. Using this data, in this paper we assess the extent to which models trained on Firefox issue discussions generalize to projects that differ in development practices, project scale, system domain, and issue management infrastructure.

The dataset includes 20 annotated issues from Chromium and GnuCash, with 10 issues per project. These projects differ in both scale and governance structure, \ie their organizational and decision-making framework. Chromium is a large desktop web browser (similar to Firefox), while GnuCash is a medium-sized desktop application for managing personal finances. Although Chromium belongs to the same application domain as Firefox, it is developed under Google’s supervision~\cite{google}, following a company-driven governance model that differs from Firefox’s community-driven approach\footnote{\url{https://www.mozilla.org/en-US/about/governance/}}\footnote{\url{https://tinyurl.com/2tnyw2cm}}. Chromium also uses a different issue-tracking infrastructure (Google Issue Tracker~\cite{google_its}) compared to Firefox’s Bugzilla-based workflow. In contrast, GnuCash is developed and maintained by volunteers and represents a smaller community-driven project from a different application domain, using the Bugzilla~\cite{bugzillaBugzilla} issue-tracking system. Together, these projects introduce variations in project size, governance structure, development practices, and issue discussion styles, offering a meaningful starting point for examining the cross-project generalizability of our models.}

\looseness=-1

To ensure issue diversity, in our prior work~\cite{saha2025decoding}, we selected the issue reports randomly following these criteria: (1) the issues were created on 1st January 2020 or later to investigate the recent issue resolution, (2) the issues are FIXED and/or RESOLVED, and (3) the issues cover different \# of comments, issue types, and creation years. 
The selected reports were analyzed and annotated using the same methodology applied to Mozilla issues (see \Cref{sub:data_col}). We then prepared the datasets for the solution identification task following the same methodology used for Mozilla data (see \Cref{sub:data_preparation}).
In total, the Chromium dataset consists of 125 comments, including 20 (16\%) solution-comments, while the GnuCash dataset contains 143 comments, including 38 (26.6\%) solution-comments (see \Cref{tab:gen_data_stats}).

\begin{table}[]
	\centering
	\caption{Generalization Dataset Statistics}
	\label{tab:gen_data_stats}
	\resizebox{\columnwidth}{!}{%
		\begin{tabular}{c|cc|cc}
			\toprule
			\textbf{Project} & \textbf{\# of Issues} & \textbf{\# of Comments} & \textbf{\# of Solution} & \textbf{\# of Non-Solution} \\ \midrule
			Chromium         & 10                    & 125                     & 20 (16.0\%)             & 105 (84.0\%)                \\
			GnuCash          & 10                    & 143                     & 38 (26.6\%)             & 105 (73.4\%)                \\ \bottomrule
		\end{tabular}%
	}
\end{table}

\subsection{Experimental Design}

\subsubsection{Evaluation Settings.}
To answer \ref{rq:gener}, we designed three complementary experimental settings. These settings help us understand how different training data sources and combinations influence model generalizability across projects.

\begin{enumerate}
    \item \textbf{Project-Specific Training}: The models are trained and evaluated on the same target project dataset (Chromium or GnuCash). This setting establishes a baseline, showing how the models perform when trained only with the limited labeled data available for a specific project, without leveraging any external knowledge.

    \item \textbf{Cross-Project Training}: The models are trained/fine-tuned on the full Mozilla dataset and evaluated on a different project (Chromium or GnuCash). This setting directly evaluates the generalization capability of the models, examining whether knowledge learned from Mozilla can transfer to projects that differ in domain, scale, and organizational structure.

    \item \textbf{Hybrid Training}: The models are trained/fine-tuned on a combined dataset that includes Mozilla and the target project data, and evaluated on the target project (Chromium or GnuCash). This setting explores whether augmenting Mozilla’s large and diverse dataset with a small amount of project-specific data can improve performance by balancing general knowledge with project-specific adaptation.
\end{enumerate}

In the \textit{Cross-Project} and \textit{Hybrid Training} settings, the full Mozilla dataset (\ie the 4,917 annotated comments---see \Cref{sub:data_preparation}) was used during fine-tuning to fully utilize the available data and maximize the transfer potential.

\subsubsection{Data Splitting Strategy.}  
One challenge in these experiments is the limited size of the new datasets, which contain annotated comments for only 10 issues per project. Chromium has an average of 12.5 comments per issue, while GnuCash contains 14.3 comments per issue. To maximize the utility of these small datasets, we adopted a leave-one-out splitting strategy at the issue report level for the \textit{Project-Specific} and \textit{Hybrid Training} settings across all models. Note that the \textit{Cross-Project Training} setting does not require data splitting, as models are trained/fine-tuned on the Mozilla dataset and evaluated on either the Chromium or GnuCash datasets.

Specifically, for each project with $n$ issues:

\begin{enumerate}
    \item The comments of the $(n-1)$ issues were used for training, while the remaining issue was held out for evaluation.
    \item This process was repeated $n$ times, each time selecting a different issue for evaluation.
    \item The final evaluation performance was computed by aggregating the \textit{true positives}, \textit{false positives}, and \textit{false negatives} across all $n$ issues.
\end{enumerate}

This strategy ensures that every available issue contributes to both training and evaluation, minimizing information loss in small datasets. Unlike traditional train-test splits, which risk discarding valuable examples, the leave-one-out approach fully leverages the limited annotated data.

\subsection{\ref{rq:gener}: Model Generalizability Results}

\begin{table}[]
\centering
\caption{Performance of the best models on Chromium and GnuCash Datasets}
\label{tab:gen_results}
\resizebox{\columnwidth}{!}{%
\begin{tabular}{cc|ccc|ccc}
\toprule
\multirow{2}{*}{\textbf{Model}}   & \multirow{2}{*}{\textbf{Setting}} & \multicolumn{3}{c|}{\textbf{Chromium}} & \multicolumn{3}{c}{\textbf{GnuCash}}  \\
                                  &                                   & \textbf{P}  & \textbf{R} & \textbf{F1} & \textbf{P} & \textbf{R} & \textbf{F1} \\ \midrule
\multirow{6}{*}{\textbf{\svcs}}     & \textbf{Project-Specific Training}               & 1.000       & 0.250      & 0.400       & 0.73       & 0.211      & 0.327       \\
                                  & \textbf{Cross-Project Training}            & 0.917       & 0.550      & 0.688       & 0.941      & 0.421      & 0.582       \\
                                  & \textbf{Hybrid Training}                   & 0.875       & 0.700      & 0.778       & 0.846      & 0.579      & 0.688       \\ \cline{2-8} 
                                  & \textbf{RI (Cross vs. Specific)}        & -8.3\%      & 120.0\%    & 71.9\%      & 29.4\%     & 100.0\%    & 78.2\%      \\
                                  & \textbf{RI (Hybrid vs. Specific)}       & -12.5\%     & 180.0\%    & 94.4\%      & 16.3\%     & 175.0\%    & 110.5\%     \\
                                  & \textbf{RI (Hybrid vs. Cross)}    & -4.5\%      & 27.3\%     & 13.1\%      & -10.1\%    & 37.5\%     & 18.2\%      \\ \midrule
\multirow{6}{*}{\textbf{\rbert}} & \textbf{Project-Specific Training}               & 0.667       & 0.500      & 0.571       & 0.750      & 0.395      & 0.517       \\
                                  & \textbf{Cross-Project Training}            & 0.722       & 0.650      & 0.684       & 0.586      & 0.447      & 0.507       \\
                                  & \textbf{Hybrid Training}                   & 0.944       & 0.850      & 0.895       & 0.949      & 0.974      & 0.961       \\ \cline{2-8} 
                                  & \textbf{RI (Cross vs. Specific)}        & 8.3\%       & 30.0\%     & 19.7\%      & -21.8\%    & 13.3\%     & -1.9\%      \\
                                  & \textbf{RI (Hybrid vs. Specific)}       & 41.7\%      & 70.0\%     & 56.6\%      & 26.5\%     & 146.7\%    & 85.8\%      \\
                                  & \textbf{RI (Hybrid vs. Cross)}    & 30.8\%      & 30.8\%     & 30.8\%      & 61.8\%     & 117.6\%    & 89.4\%      \\ \midrule
\multirow{6}{*}{\textbf{\llamaft}}   & \textbf{Project-Specific Training}               & 0.120       & 0.300      & 0.171       & 0.386      & 0.711      & 0.500       \\
                                  & \textbf{Cross-Project Training}            & 0.438       & 0.350      & 0.389       & 0.524      & 0.579      & 0.550       \\
                                  & \textbf{Hybrid Training}                   & 0.500       & 0.500      & 0.500       & 0.491      & 0.737      & 0.589       \\ \cline{2-8} 
                                  & \textbf{RI (Cross vs. Specific)}        & 264.6\%     & 16.7\%     & 126.9\%     & 35.8\%     & -18.5\%    & 10.0\%      \\
                                  & \textbf{RI (Hybrid vs. Specific)}       & 316.7\%     & 66.7\%     & 191.7\%     & 27.4\%     & 3.7\%      & 17.9\%      \\
                                  & \textbf{RI (Hybrid vs. Cross)}    & 14.3\%      & 42.9\%     & 28.6\%      & -6.2\%     & 27.3\%     & 7.2\%       \\ \bottomrule
\end{tabular}%
}
\end{table}

\Cref{tab:gen_results} presents the performance of the three best models under the three training settings on both Chromium and GnuCash datasets. Overall, the results show that models fine-tuned on Mozilla data can generalize to other projects, and incorporating a small amount of project-specific data further enhances their performance. Across all models and both projects, we observed consistent trends across the three settings. The \textit{Project-Specific Training} setting resulted in the lowest performance, with average F1 scores of $0.381$ and $0.448$ across three models for Chromium and GnuCash, respectively, suggesting that limited labeled data within a single project is insufficient for learning robust patterns to identify solution-related comments. 

In contrast, the \textit{Cross-Project Training} setting---where models fine-tuned only on Mozilla data were evaluated on other projects---achieved higher average F1 scores of $0.587$ (Chromium) and $0.546$ (GnuCash), improving over \textit{Project-Specific Training} by $54.1\%$ and $22\%$, respectively. This finding confirms that models fine-tuned on Mozilla data can transfer useful knowledge to other projects, even without additional project-specific fine-tuning. 

However, the best performance was achieved in the \textit{Hybrid Training} setting, where Mozilla and project-specific data were utilized during fine-tuning. This setup yielded the highest average F1 scores of $0.724$ (Chromium) and $0.746$ (GnuCash), representing gains of $90.1\%$ and $66.5\%$ over \textit{Project-Specific Training} and $23.4\%$ and $36.5\%$ over \textit{Cross-Project Training}. These results demonstrate that while models trained on Mozilla data alone exhibit cross-project generalizability, combining them with even a small amount of target project data provides the best balance between generalization and adaptation.

Interestingly, while \llamaft achieved the best performance on the larger Mozilla dataset (4,917 comments, including 808 solution comments) in our main experiments (\Cref{sec:results}), \rbert and \svcs performed better in the generalization study on smaller datasets—Chromium and GnuCash—with 125 and 143 comments (20 and 38 solution comments), respectively. Both models outperformed \llamaft under the \textit{Cross-Project} and \textit{Hybrid Training} settings for both datasets, except for \rbert in the \textit{Cross-Project Training} setting on GnuCash. These results suggest that larger LLMs such as \llamaft may require substantially more project-specific examples to adapt effectively across domains. This observation is consistent with prior research showing that fine-tuned LLMs often struggle when labeled target-domain data is limited~\cite{jeong2024fine,lu2025fine}. In practice, for new projects with limited annotation data, a fine-tuned PLM like \rbert—or even a lightweight model like \svcs—may offer greater robustness and reliability than a heavily parameterized LLM, unless sufficient in-project data is available for adaptation.

\subsubsection{Cross-Project vs. Project-Specific Training.}  
Comparing the \textit{Cross-Project} and \textit{Project-Specific Training} settings shows that models fine-tuned on Mozilla data can generalize to other projects, often outperforming models trained only on limited project-specific data. Across all three models and both projects, we observed clear improvements in F1 scores under the \textit{Cross-Project Training} setting, indicating that the knowledge learned from Mozilla transfers well across different projects.

For \svcs, \textit{Cross-Project Training} substantially improved performance, with F1 scores increasing from $0.400$ to $0.688$ ($+71.9\%$) on Chromium and from $0.327$ to $0.582$ ($+78.2\%$) on GnuCash. These gains were driven primarily by higher recall, while precision remained consistently high (avg.\ $\approx 0.92$). This suggests that traditional feature-based models like \svcs benefit greatly from the large and diverse Mozilla dataset, which enables them to detect a wider range of solution-related comments without sacrificing precision.

For \rbert, results were mixed. On Chromium, \textit{Cross-Project Training} improved F1 from $0.571$ to $0.684$ ($+19.7\%$), with modest gains in both precision ($0.667 \rightarrow 0.722$) and recall ($0.5 \rightarrow 0.65$). However, on GnuCash, performance slightly decreased ($0.517 \rightarrow 0.507$, $-1.9\%$) due to lower precision ($0.75 \rightarrow 0.586$) despite improved recall ($0.395 \rightarrow 0.447$). This indicates that \rbert fine-tuned on Mozilla data generalizes well to projects with similar characteristics (like Chromium), but less effectively to projects that differ substantially in domain and scale (like GnuCash).

For \llamaft, \textit{Cross-Project Training} also led to consistent gains. On Chromium, F1 improved from $0.171$ to $0.389$ ($+126.9\%$), and on GnuCash from $0.500$ to $0.550$ ($+10.0\%$), primarily due to better precision (Chromium: $0.12 \rightarrow 0.438$; GnuCash: $0.386 \rightarrow 0.524$). These improvements show that fine-tuning \llama on Mozilla enables it to generalize to new projects with different domains and structures, capturing more abstract patterns of solution-related language.

Overall, models fine-tuned on Mozilla data can generalize well to other projects. The improvement is more significant for models that rely heavily on data diversity (like \svcs), while more advanced language models (\rbert and \llamaft) exhibit varying degrees of transfer depending on the similarity between the source and target projects.

\subsubsection{Hybrid vs. Project-Specific and Cross-Project Training.}  
The \textit{Hybrid Training} setting, which combines Mozilla data with the small project-specific datasets, consistently produced the best results across all models and projects. Overall, these results indicate that while models fine-tuned on Mozilla data can generalize effectively, incorporating even a small amount of project-specific data further enhances performance by enabling better adaptation to local project characteristics.

For \svcs, \textit{Hybrid Training} achieved the highest F1 scores of $0.778$ and $0.688$ on Chromium and GnuCash, respectively—improving over \textit{Project-Specific Training} by $94.5\%/110.5\%$ and over \textit{Cross-Project Training} by $13.1\%/18.2\%$. These gains were driven by concurrent improvements in both precision and recall, suggesting that while \svcs can learn general solution-related patterns from Mozilla data, it still relies on project-specific examples to refine those patterns and reach optimal performance.

For \rbert, a similar trend emerged, with \textit{Hybrid Training} achieving the highest overall F1 scores among all models—$0.895$ on Chromium and $0.961$ on GnuCash. These represent improvements of $56.6\%/85.8\%$ over \textit{Project-Specific Training} and $30.8\%/89.4\%$ over \textit{Cross-Project Training}. The model achieved balanced gains in both precision and recall, demonstrating that integrating project-specific data enables \rbert to adapt effectively while retaining the generalizable knowledge learned from Mozilla. This combination makes \rbert the most reliable model for cross-project solution comment identification. The superior performance is aligned with observations that encoder‑only architectures such as \rbert provide robust representations for downstream classification and generalize well across tasks after fine‑tuning~\cite{liu2019roberta}.

\llamaft also benefited from \textit{Hybrid Training}, with F1 scores increasing from $0.171/0.389$ to $0.500$ on Chromium and from $0.500/0.550$ to $0.589$ on GnuCash compared to the \textit{Project-Specific} and \textit{Cross-Project} settings. Although the absolute gains were smaller than those of \rbert, the results show that LLM-based models can effectively combine general knowledge from large-scale data with minimal project-specific supervision, further improving precision and recall. The relatively lower performance of \llamaft compared to \rbert is consistent with recent findings that decoder‑only architectures often underperform encoder models in sequence‑labeling or classification tasks under limited fine‑tuning scenarios~\cite{dukic2024looking}.

In summary, \textit{Hybrid Training} clearly outperforms both \textit{Project-Specific} and \textit{Cross-Project} training. Combining the large and diverse Mozilla dataset with even limited project-specific annotations provides the best balance between generalization and adaptation, yielding strong and consistent performance across projects and models. These findings confirm that a small amount of in-domain data is critical to unlocking the full transfer potential of models fine-tuned on Mozilla data.

\rqanswer{\textbf{\ref{rq:gener} Findings}: The best models from each type (\ie\ \svcs, \rbert, and \llamaft) fine-tuned on Mozilla data generalize well to other projects, demonstrating strong cross-project transferability. The \textit{Cross-Project Training} setting outperforms \textit{Project-Specific Training}, showing that knowledge from Mozilla captures reusable patterns of solution-related comments. The best results, however, come from the \textit{Hybrid Training} setting, where combining Mozilla data with limited project-specific examples yields the highest precision and recall. Among models, \rbert shows the strongest adaptability, while \svcs and \llamaft also benefit notably from the hybrid approach. These findings highlight that Mozilla-based fine-tuning provides a solid foundation for generalization, and even small project-specific datasets further enhance performance.}

\section{Threats to Validity}

\textbf{Construct and Internal Validity.} 
Researcher subjectivity and potential confirmation bias during qualitative analysis and result interpretation pose key threats to construct and internal validity. We mitigated these risks through a rigorous and systematic open-coding approach in our qualitative study. Each comment was independently analyzed by two researchers, followed by a consensus discussion to finalize the reasoning behind each model’s failure. To assess the reliability of the coding process, we measured inter-coder agreement (see \Cref{sub:qual_method}), which showed a high level of consistency between coders. The interpretation of results involved multiple researchers and was carefully reviewed to ensure consistency and support from data-driven evidence.

The finetuning and hyperparameter selection for each model, as well as the selection of the model, could impact the results. While we conducted hyperparameter tuning for the machine learning models (MLMs) and experimented with different batch sizes and epochs for PLMs and LLMs, the chosen parameters may not be optimal. Due to resource limitations, we were unable to explore a wider range of batch sizes and other hyperparameters, which could potentially lead to suboptimal model performance and incomplete conclusions about their true capabilities.

LLM data leakage may pose another threat to validity. It is possible that \llama was trained on large amounts of web data, potentially including the Firefox/Chromium/GnuCash issues used for testing. However, this is likely not a significant concern, as \llama performed suboptimal with prompting, showing lower performance than both MLMs and PLMs. In contrast, the fine-tuned version of \llama exhibited substantially better results and emerged as the best-performing individual model.

\textbf{External Validity.} 
As is typical for empirical studies, our findings may not generalize to all software systems, domains, or issue-tracking environments. To mitigate this threat, we conducted a generalizability study (see \Cref{sec:generalizability}) to evaluate whether the models trained/fine-tuned on Mozilla data can be effectively applied to other projects with different scales, domains, and governance structures (\ie organizational and decision-making framework). Specifically, we tested the best-performing models on issue reports from Chromium and GnuCash—two projects that differ substantially from Mozilla in terms of size, community, and issue-tracking systems. 

The results demonstrated that while the models trained solely on Mozilla data achieved reasonable cross-project performance, combining large-scale Mozilla data with even limited project-specific examples (\ie the \textit{Hybrid} configuration) yielded the most robust and consistent results across both projects. These findings suggest that the learned patterns are not overfitted to Mozilla and can transfer to other software contexts, particularly when minimal target-project data is available for fine-tuning. Nonetheless, the generalizability of our conclusions remains constrained by the limited number of external projects and annotated issues used in this study. Future work should replicate and extend this investigation across a broader set of projects, domains, and issue-tracking systems to strengthen the external validity of our results.

\section{Discussion and Implications}
\label{sec:implications}

In this section, we discuss the key takeaways from our findings and outline their practical implications for researchers and practitioners for automating solution-related content identification in issue discussions.

\indent \textbf{Automating Solution Identification in Issue Discussions.}
Manually analyzing long issue discussions to locate solution-related content is both time-consuming and cognitively demanding. Our work on automatically identifying solution-related comments can meaningfully streamline this process. The best-performing fine-tuned model (\llamaft) achieves an F1 score of 0.716 with a recall of 0.838, suggesting that it can identify most true solution suggestions while keeping noise manageable. In practice, such models can be integrated into issue tracking systems to highlight or summarize comments likely containing solutions, helping developers quickly revisit past resolutions, debug regressions, or reuse fixes. Beyond individual developer productivity, this automation can also support project-level analytics—such as tracking common solution patterns or assessing how quickly and effectively teams respond to and resolve issues—thereby improving overall software maintenance efficiency. However, it remains unclear how practical these performance levels are in real development settings. Future work can conduct user studies to examine how such automated support fits into developers’ workflows and whether the achieved precision and recall actually lead to meaningful productivity improvements.

\rev{\indent \textbf{Context-Aware Solution Identification.}
Our qualitative analysis shows that even the best-performing models (\ie\ \svc, \rbert, and \llamaft) often struggle when comments contain misleading clues (\eg keywords, code snippets, or technical jargon) or lack sufficient contextual information (\eg the issue description or surrounding discussion) needed to interpret intent accurately. In this study, we deliberately analyze each comment independently to evaluate how well models can identify solution-related content based solely on the information contained in that comment. This design allows us to isolate the intrinsic capability of models to recognize solution-related content without relying on additional contextual information. (We note though, that comments still may provide some contextual information as they can contain content of different kinds, not just solution-related content.)

However, issue discussions typically evolve as a conversation, where earlier comments describe the problem or unexpected behavior and later comments may propose potential fixes or workarounds. Ignoring such discussion history can make it difficult to interpret short or ambiguous comments and may contribute to the performance ceiling observed in our results. Incorporating richer contextual information—such as the issue description, preceding comments, or the overall discussion structure—therefore represents an important direction for future work. Approaches based on hierarchical, conversational, or graph-based modeling could leverage these dependencies to better capture how solutions emerge throughout the discussion.

Relatedly, our LLM-prompting results suggest that including the full issue description as context may sometimes introduce excessive or irrelevant information for large language models. Future work could explore methods for identifying and extracting the most salient parts of issue descriptions—such as key problem statements, error messages, or reproduction steps—to provide more focused contextual inputs that improve classification accuracy.}

\rev{\textbf{Performance and Implementation Effort Trade-Off.}
Our results reveal a clear trade-off between predictive performance and implementation effort across different modeling strategies. Traditional ML classifiers combined with modern language model embeddings can achieve competitive performance while remaining computationally lightweight. For example, \svc with \embgpt embeddings performs comparably to several PLMs and LLM while relying on a substantially simpler training and deployment pipeline. Such approaches allow teams with limited computational resources to still benefit from the semantic richness of pretrained embeddings.

At the same time, task-specific fine-tuning consistently yields the best overall performance. The fine-tuned \llamaft and \rbert models substantially outperform both traditional ML models and prompted LLMs, achieving F1/recall scores of 0.716/0.838 and 0.692/0.712, respectively. Fine-tuning enables models to learn the linguistic and contextual patterns of solution discussions—something prompting alone struggles to capture. However, prompting-based approaches remain attractive because they require minimal model adaptation and can be applied directly to pretrained LLMs.

Taken together, these findings suggest that different strategies may be appropriate depending on the available resources, required accuracy, and deployment constraints. Lightweight embedding-based classifiers may provide an efficient baseline solution, while fine-tuned models offer the highest performance when sufficient labeled data and computational resources are available. A more systematic investigation of this trade-off—considering factors such as computational cost, training time, and maintenance effort alongside predictive performance—remains an important direction for future work and could help practitioners make more informed decisions when selecting modeling strategies for real-world adoption.}

\textbf{Cross-Project Adaptation and Transferability.}
Our \textit{Cross-Project Training} experiments provide evidence that solution-comment identification models can generalize across projects. Models fine-tuned on the large and diverse Mozilla dataset perform reasonably well on other projects such as Chromium and GnuCash, and their performance improves further when supplemented with a small amount of target-project data. The \textit{Hybrid Training} setup—combining Mozilla data with limited project-specific examples—consistently achieved the best precision and recall, indicating that even modest local adaptation helps reduce domain gaps in terminology, coding conventions, and discussion style.

Interestingly, while \llamaft achieved the best performance on the larger Mozilla dataset, \rbert and \svcs performed better in the generalization study on the smaller Chromium and GnuCash datasets. These results suggest that larger LLMs such as \llamaft may require substantially more project-specific examples to adapt effectively across domains, whereas smaller PLMs or lightweight models can generalize more robustly when only limited labeled data is available. In practice, developers can start from our publicly available models and datasets, then refine them with only a handful of annotated issues from their own projects to achieve strong performance. This lightweight adaptation strategy offers a practical and scalable way to deploy solution identification tools across different projects.

\rev{\textbf{Towards Practical and Trustworthy Solution Identification.}
While our study demonstrates the feasibility of automated solution identification, further research is needed to better understand how such tools can be used effectively in practice. For example, developers may benefit from explanations or rationales accompanying model predictions to help them assess why a comment is identified as solution-related. Incorporating explainable AI techniques or evidence-based highlighting within comments could improve transparency and foster developer trust in automated recommendations. In addition, user-centered evaluations—such as developer studies or field deployments—would help assess whether these tools meaningfully support debugging, issue resolution, and knowledge reuse in real-world workflows.}

\rev{\textbf{Advancing Modeling Techniques.}}
Finally, there remains significant opportunity to further improve model performance. The ensemble of the best models from each type achieves an F1 of 0.737, highlighting both the difficulty of the task and the potential for further advancement. Future research could explore more sophisticated ensembles or architectures that jointly leverage the complementary strengths of MLMs, PLMs, and LLMs. Another promising avenue is to develop human-in-the-loop systems that combine automated identification with developer validation, enabling users to confirm or refine predictions interactively. Such feedback-driven systems could improve performance over time while fostering trust and adoption in real-world development workflows.

\section{Conclusion}
\label{sec:conclusions}

To assist developers in identifying solution-related content in issue report discussions, we conducted a case study evaluating 12 language model-based text classifiers. This study analyzed 4,917 comments from 356 Mozilla Firefox issue reports. We tested six traditional machine learning models (MLMs), four pre-trained language models (PLMs), and two large language models (LLMs) across 68 different configurations, focusing on three approaches: embeddings, prompting, and fine-tuning. We also evaluated combinations of the best models using a majority voting approach to assess their collective predictive power.

Our study revealed several key insights into the effectiveness of different classifiers. MLMs are more effective when using LM embeddings than \embtfidf for encoding issue comments, showcasing LMs' superior ability to capture text semantics for solution comment classification. \svc with \embgpt achieved the highest MLM performance without data balancing, reaching an F1 score of 0.663. In terms of prompted classifiers, \llama with chain-of-thought (CoT) prompting underperformed, with the lowest F1 score of 0.517. However, fine-tuned LMs showed the best results overall, with fine-tuned \llama leading at 0.716 F1, 0.626 precision, and 0.838 recall, followed by \rbert at 0.692 F1, 0.674 precision, and 0.712 recall. Among ensemble models, combining the best-performing models from each type—\svcs, \rbert, and \llama—yielded the highest performance with an F1 score of 0.737, precision of 0.702, and recall of 0.775, highlighting the effectiveness of leveraging diverse models. 

Our qualitative analysis showed that models often misclassified comments when they include misleading clues (\eg technical jargon, keywords, code snippets, \etc) or lack contextual information (\eg issue description and surrounding discussion). The generalizability study confirmed that limited project-specific data alone is insufficient, while cross-project transfer from Mozilla improved performance. The best results came from a hybrid approach combining Mozilla and project-specific data, highlighting the importance of both large external datasets and in-project examples for accurately identifying solution-related content across projects.

\section{Declarations}

\subsection{Funding}
This work is supported in part by the National Science Foundation (NSF) grants: CCF-2239107 and CCF-1955853. Any opinions, findings, and conclusions expressed herein are the authors and do not necessarily reflect those of the sponsors.

\subsection{Ethical Approval}
This study does not involve experiments with human participants or animals. The research is based solely on the analysis of publicly available data from open-source software repositories and issue tracking systems. Therefore, ethical approval from an institutional review board or ethics committee was not required.

\subsection{Informed Consent}
This study did not involve human participants, and no personal or identifiable data were collected. Therefore, informed consent was not required.

\subsection{Author Contributions}
All authors contributed to the conception and design of the study. Material preparation, data collection, experimental design, implementation, execution, and analysis were performed by Antu Saha and Mehedi Sun. Oscar Chaparro provided supervision and mentorship throughout the project and contributed to the design of the data collection process, experimental methodology, and analysis of results through review and feedback.

The first draft of the manuscript was written by Antu Saha, and all authors reviewed, edited, and commented on previous versions of the manuscript. All authors read and approved the final manuscript.

\subsection{Data Availability Statement}
For verifiability and reproducibility, we have made all the study artifacts (\eg\ dataset, source code, and documentation) publicly available \cite{repl_pack}.

\subsection{Conflict of Interest}
The authors declare that they have no conflict of interest.

\subsection{Clinical Trial Number}
Not applicable.

\bibliographystyle{spmpsci}      %
\bibliography{references}   %

\end{document}